\documentclass[preprint,12pt]{elsarticle}

\usepackage{amssymb,amsmath,sectsty}
\usepackage{color,graphicx,epsfig}
\usepackage{rotating}
\usepackage{subfigure}
\usepackage{dcolumn}
\usepackage{bm}
\usepackage{enumerate}

\newcommand{\ds}{\,\text{d}s}
\newcommand{\dt}{\,\text{d}t}

\newcommand{\dtheta}{\,\text{d}\theta}
\newcommand{\deta}{\,\text{d}\eta}

\newcommand{\bfd}{\boldsymbol{d}}
\newcommand{\bft}{\boldsymbol{t}}
\newcommand{\bfe}{\boldsymbol{e}}
\newcommand{\bfx}{\boldsymbol{x}}
\newcommand{\bfn}{\boldsymbol{n}}

\newcommand{\bfp}{\boldsymbol{p}}
\newcommand{\bfr}{\boldsymbol{r}}

\newcommand{\eps}{\varepsilon}

\newcommand{\bfkn}{\boldsymbol{\kappa_n}}
\newcommand{\bfkg}{\boldsymbol{\kappa_g}}
\newcommand{\bfk}{\boldsymbol{\kappa}}

\newcommand{\kn}{\kappa_n}
\newcommand{\kg}{\kappa_g}

\newcommand{\abs}[1]{|#1\rvert}

\newcommand{\etal}{\it{\text{et al.}}}

\newfont{\tenbss}{bbmss12}

\newfont{\tenbfsl}{cmbxti12}

\newcommand{\pa}{\partial}

\newcommand{\ct}{\cos\theta}
\newcommand{\ce}{\cos\eta}
\newcommand{\st}{\sin\theta}
\newcommand{\se}{\sin\eta}

\DeclareMathOperator\erf{erf}

\journal{International Journal of Non-linear Mechanics}

\begin{document}

\begin{frontmatter}



{\author{Meisam Asgari\corref{cor1}}
\ead{meisam.asgari@mail.mcgill.ca}
\cortext[cor1]{Corresponding author}}

\author{Aisa Biria\corref{}}
\address{Department of Mechanical Engineering, McGill University, Montr\'eal, QC H3A 0C3, Canada}

\title{Free energy of the edge of an open lipid bilayer based on the interactions of its constituent molecules}

\begin{abstract}
Lipid-bilayers are the fundamental constituents of the walls of most living cells and lipid vesicles, giving them shape and compartment. The formation and growing of pores in a lipid bilayer have attracted considerable attention from an energetic point of view in recent years. Such pores permit targeted delivery of drugs and genes to the cell, and regulate the concentration of various molecules within the cell. The formation of such pores is caused by various reasons such as changes in cell environment, mechanical stress or thermal fluctuations. Understanding the energy and elastic behaviour of a lipid-bilayer edge is crucial for controlling the formation and growth of such pores. In the present work, the interactions in the molecular level are used to obtain the free energy of the edge of an open lipid bilayer. The resulted free-energy density includes terms associated with flexural and torsional energies of the edge, in addition to a line-tension contribution. The line tension, elastic moduli, and spontaneous normal and geodesic curvatures of the edge are obtained as functions of molecular distribution, molecular dimensions, cutoff distance, and the interaction strength. These parameters are further analyzed by implementing a soft-core interaction potential in the microphysical model. The dependence of the elastic free-energy of the edge to the size of the pore is reinvestigated through an illustrative example, and the results are found to be in agreement with the previous observations.

\end{abstract}

%

\end{frontmatter}


\section{Introduction}
\label{Introduction}

A phospholipid molecule consists of a hydrophilic head and two hydrophobic fatty-acid tails~\cite{israelachvili2011intermolecular}. When suspended in an aqueous solution at sufficient concentrations, phospholipid molecules self-assemble into structures such as lipid bilayers, in order to shield the tail groups from the solvent~\cite{israelachvili1977theory,israelachvili1976theory}. Lipid bilayers are the main constituents of cell membrane in most living organisms, as well as model membranes such as liposomes~\cite{lodish2000molecular}. They provide the cell and its substructures with compartment and shape, and further, function as barriers for water-soluble molecules such as water, ions, and proteins~\cite{saitoh1998opening,luckey2014membrane}. Lipid bilayers are composed of two adjacent leaflets of phospholipid molecules oriented transversely and set tail-to-tail. 

Forming of open edges in lipid membranes results in exposer of the tail groups at the edge to water~\cite{lodish2000molecular}, which is energetically unfavourable. As a result, phospholipid molecules rapidly rearrange around the exposed edge, forming a semicylindrical rim along it. This rearrangement is the source of a line energy at the edge. In order to eliminate this edge energy, lipid bilayers commonly tend to form closed structures such as spheroids~\cite{seifert1997configurations}. Nevertheless, they can transiently open due to various stimuli such as mechanical stresses and thermal instabilities. The formation of these transient pores is essential for regulation of PH, transmembrane electrochemical potential, and concentrations of different molecules in the cell~\cite{saitoh1998opening}. Additionally, transient open membranes are formed during electro-formation~\cite{lasic1988mechanism}. More recently, stabilizing pores and control over their size have been pursued by means of electric fields~\cite{tsong1991electroporation}, sonication~\cite{marmottant2003controlled}, and use of edge-active chemical agents~\cite{fromherz1986discoid}. The rapid progress in these techniques has attracted increasing attention to the study of the open lipid bilayers, including molecular dynamic simulations, as well as continuum mechanical treatment and numerical investigations of the equilibrium configurations~\cite{tu2004geometric,tu2008elastic}.

Theoretical studies of the equilibrium and stability of pored membranes have mainly relied on constitutive assumptions for the edge, which neglect its flexural and torsional elasticity. For instance, Boal and Rao~\cite{boal1992scaling}, Capovilla {\etal}~\cite{capovilla2002lipid}, and Tu and Ou-Yang~\cite{tu2010compatibility,tu2011geometry} considered the edge energy of an open lipid bilayer as a given constant. Tu and Ou-Yang~\cite{tu2003lipid} considered dependence of the edge energy on its geometry, namely geodesic and normal curvatures. Nevertheless, their assumptions on the form of the line energy have not been precisely justified.

May~\cite{may2000molecular} obtained the line energy of a lipid bilayer edge through optimization of the lipid packing at the vicinity of the edge. He modeled the edge as a semicylindrical micelle, and took the free energy per molecule to depend upon the chain length of the molecules, their cross-sectional area, and the strength of the interactions of the molecules with each other and with the surrounding solution. Although successful in obtaining the line tension, that framework did not capture the bending and torsional energetics of the edge. The gap in the literature to successfully relate the macro-scale edge energy to its microstructure has motivated the current study.

The interactions between the constituent molecules of a material may be used to obtain the free-energy density function of that material. For instance, Keller and Merchant~\cite{keller1991flexural} have employed such a microphysical approach to extract the internal energy, surface tension, and bending energy of a liquid surface and to relate its bending rigidity to the molecular density and interaction potential. In a recent application of the work of Keller and Merchant~\cite{keller1991flexural}, the Canham--Helfrich free-energy density for a lipid vesicle was derived based on microphysical considerations~\cite{seguin2014microphysical}. Using the same approach, a model for the elastic free-energy of wormlike micelles was derived~\cite{asgari2015elastic}. In so doing, the surfactant molecules comprising the wormlike micelle were assumed to have constant length, and thus, were modeled by one-dimensional rigid rods. The resulted expression for the free energy was found to be quadratic in the curvature and torsion of the centerline of the micelle~\cite{asgari2015elastic}.
 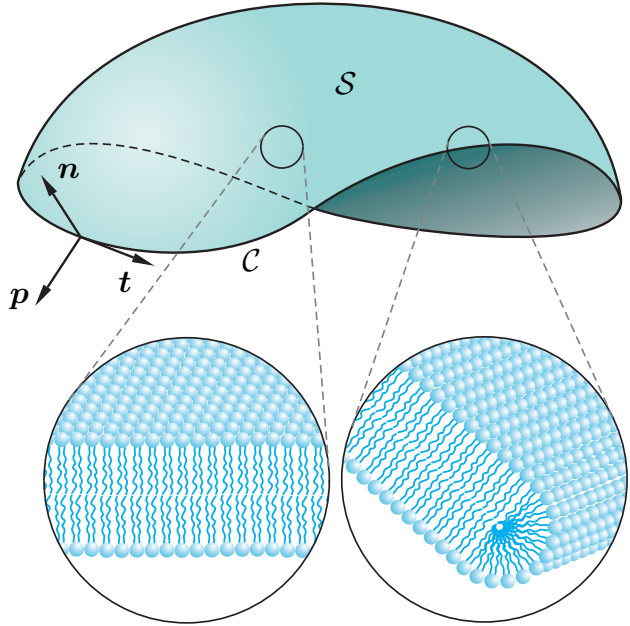
\begin{figure} [!t]
  \centering
  \includegraphics [height=3.4in] {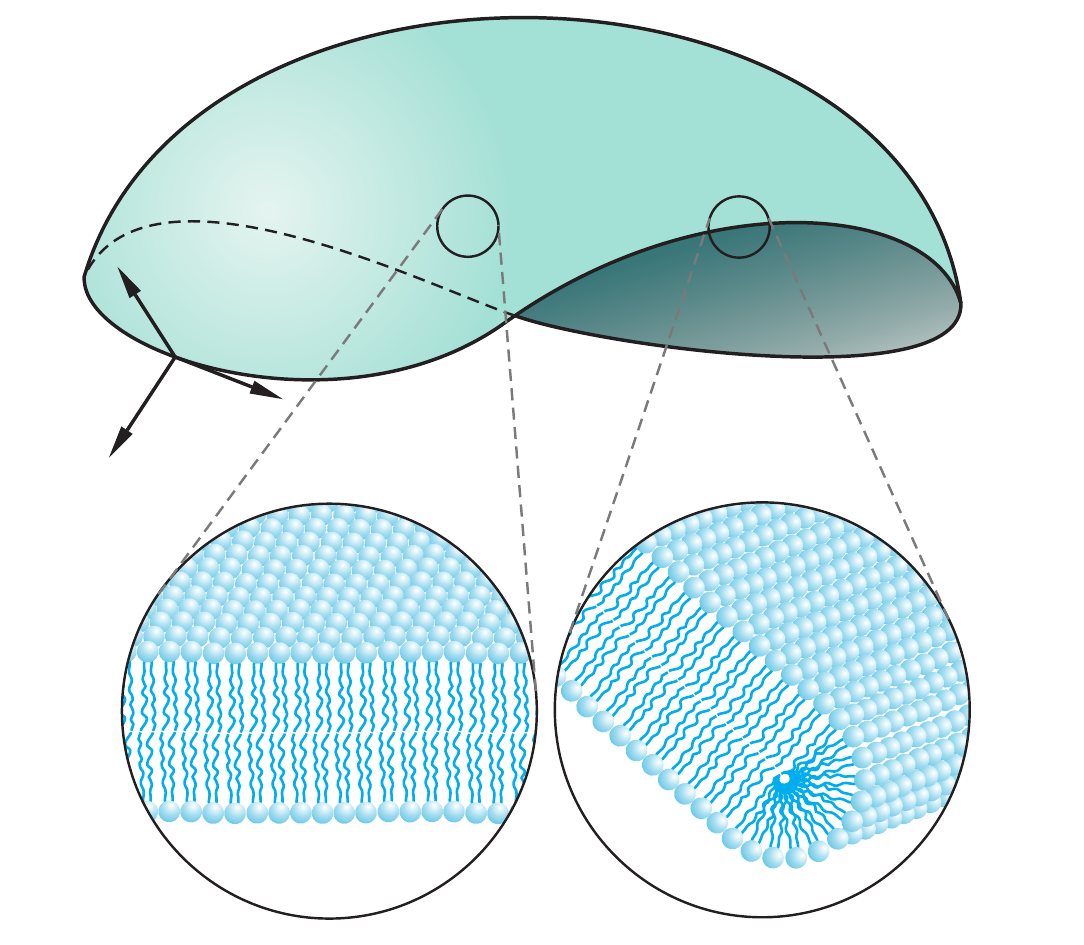}
  \put(-140,207){$\cal S$}   
  \put(-175,140){$\cal C$}
  \put(-245,175){$\bfn$}
  \put(-222,132){$\bft$}
  \put(-263,128){$\bfp$}
\caption{Mathematical identification of an open lipid bilayer as an open surface $\cal S$ with boundary $\cal C=\pa \cal S$ on which a Darboux frame has been shown. Also the schematic arrangements of phospholipid molecules in an interior point on $\cal S$ and at the vicinity of the edge $\cal C$ are depicted at a point.}
\label{Fig1}
\end{figure}

The current study adopts the microphysical approach of Keller and Merchant~\cite{keller1991flexural} to investigate the elastic behaviour of the edge of a lipid bilayer. Following May~\cite{may2000molecular} and motivated by previous studies~\cite{jiang2004molecular,wohlert2006free,de2006coarse,karatekin2003cascades}, the edge is modeled as a semicylindrical surface. In addition, the phospholipid molecules comprising the edge are modeled as one-dimensional rigid rods of constant length, oriented perpendicular to the centerline of the edge. The applied framework enables us to extract the form of the free energy and the flexural and torsional moduli of the edge, based on the intermolecular energetic interaction between phospholipid molecules.

To find the free-energy density of the edge at a position $\bfx$, we account for the interactions between all phospholipid molecules on the edge within a cutoff distance $\delta$ from the molecules at $\bfx$. We assume that the phospholipid molecules are perpendicular to the centerline of the edge. Our derivation relies on Taylor series expansions with respect to a dimensionless parameter $\varrho:=\delta/\ell\ll1$, where $\ell$ is a characteristic size parameter of the edge, such as its length. For $\ell$ taken as the length of the edge (or equivalently, the perimeter of a pore), it can be related to the thickness or the length of the constituent molecules, if the density of the molecules along the edge and their aspect ratios are provided. The net free-energy of the edge results from integrating the free-energy density $\phi$ over the centerline of the edge. 

The paper is structured as follows. In Section~\ref{Differential geometry}, required mathematical definitions are presented. Modeling assumptions for the edge of an open lipid bilayer are synopsized in Section~\ref{assumptions}. Section~\ref{derivation} is concerned with the derivation of the free-energy density of such an edge. In Section~\ref{results}, the consequences of choosing a spheroidal-particle potential (Berne and Pechukas~\cite{berne1972gaussian} and Gay and Berne~\cite{gay1981modification}) are considered to obtain the material parameters present in the derived model. As an illustrative example, a simplified model for a pore on a lipid bilayer is given in Section~\ref{Illustrative}, and the parameters obtained in Section~\ref{results} are used to find the free-energy of the pore as a function of its size. Finally, the key findings of the study are summarized and discussed in Section~\ref{discussion}. Details of the various derivations are provided in the Appendix.

\section{Differential geometry of the bounding curve of a surface}\label{Differential geometry}

Consider a smooth, orientable, open surface $\cal S$ representing the open lipid bilayer, with boundary $\cal C=\pa {\cal S}$, as depicted schematically in Figure~\ref{Fig1}. Let 
\begin{equation}
\label{curve}
{\cal C}=\{\bfx:\bfx=\bfx(s),0\le s\le L\},
\end{equation}
denote the arclength parametrization of the closed boundary curve $\cal C$. On denoting the differentiation with respect to the arclength $s$ by a superposed dot, 
it follows that $\abs{\dot\bfx}=1$, and thus, 
\begin{equation}
\label{constraint2}
\dot\bfx\cdot\ddot\bfx=0,\quad\text{and}\quad\abs{\dot\bfx\times\ddot\bfx}=\abs{\ddot\bfx}. 
\end{equation}
The unit tangent of $\cal C$ is introduced, in terms of the arclength parametrization $\bfx$, by
\begin{equation}
{\bf{t}}:=\dot{\bfx}.
\end{equation}
Since the unit tangent $\bf{t}$ has a constant length, its arclength derivative $\dot{\bf{t}}=\text{d}{\bf{t}}/\ds$ is perpendicular to it, and thus, perpendicular to the curve $\cal C$. The orientation of $\dot{\bf{t}}$ is called the unit normal of $\cal C$, and is denoted by $\bf{N}$. The curvature vector $\bfk$ at any point of $\cal C$ is then defined by the arclength derivative of the unit tangent $\bf{t}$ as
\begin{equation}
\bfk:=\dot{\bf{t}}
=\kappa\,\bf{N},
\label{b2}
\end{equation}
where $\kappa$ denotes the magnitude of the curvature of $\cal C$ at that point, which is given in terms of the arclength parametrization $\bfx$, by
\begin{equation}
\kappa={\abs{\dot{\bfx}\times\ddot{\bfx}}}=\abs{\ddot{\bfx}}.
\label{curvature8}
\end{equation}
For an arbitrary point on curve $\cal C$ at which $\kappa\ne0$, the unit binormal vector is defined by $\bf{B}=\bf{t}\times\bf{N}$. The unit tangent $\bf{t}$, unit normal $\bf{N}$, and unit binormal $\bf{B}$ at each point of $\cal C$, form the Frenet frame $\{\bf{t},\bf{N},\bf{B}\}$ at that point.

The torsion $\tau$ of $\cal C$ is defined by $\dot{\bf{B}}=-\tau\bf{N}$, and is expressed in terms of the arclength parametrization ${\bfx}$ as
\begin{equation}
\tau=\frac{\dot{\bfx}\cdot(\ddot{\bfx}\times\dddot{\bfx})}{\abs{\ddot{\bfx}}^2}.
\label{curvtor}
\end{equation}
The torsion $\tau$ of $\cal C$, describes the tendency of the curve $\cal C$ to move out of its osculating plane at a given point, or, equivalently, it measures the turnaround of the unit binormal $\bf{B}$ of $\cal C$ at a given point. In general, a space curve is determined up to a rigid translation, by its two locally invariant quantities: the curvature $\kappa$ and torsion $\tau$, both in terms of the arclength parameter $s$. 
 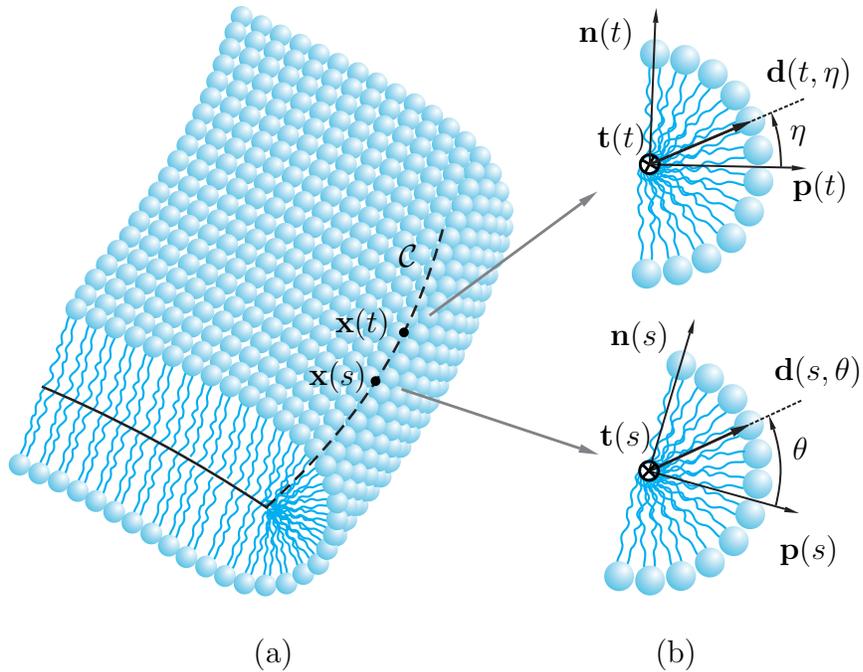
\begin{figure} [t]
  \centering
  \includegraphics [height=3.1in] {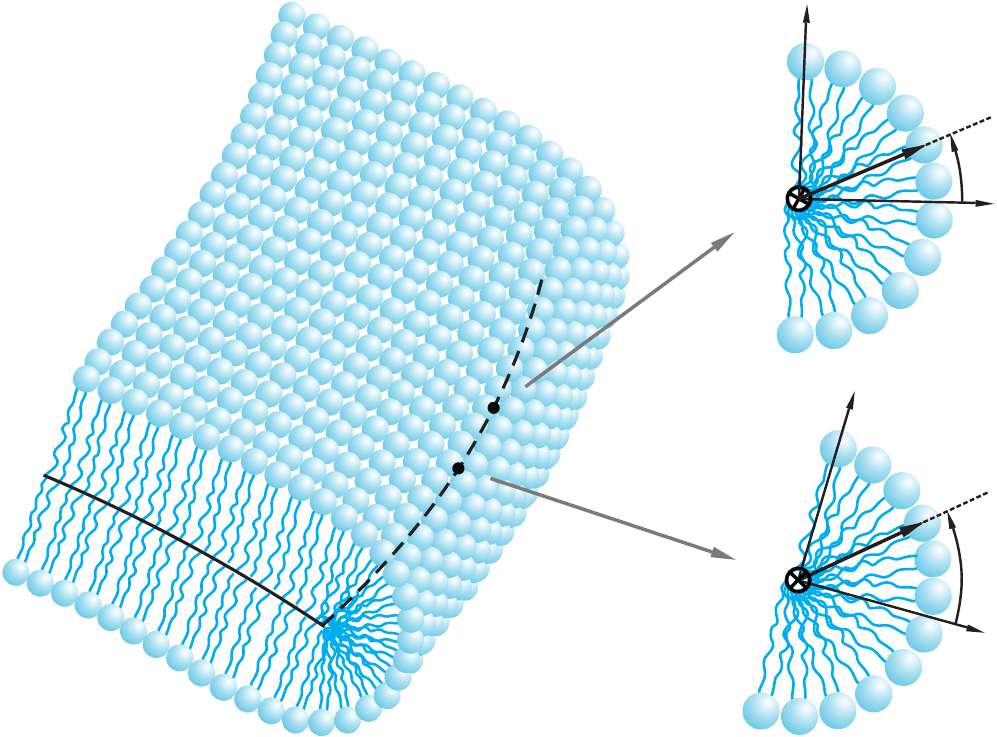}
  \put(-210,-25){$\text{(a)}$}   
  \put(-58,-25){$\text{(b)}$}   
  \put(-155,125){$\cal C$}   
  \put(-179,100){${\bf{x}}(t)$}   
  \put(-189,80){${\bf{x}}(s)$}   
  \put(-87,210){${\bf{n}}(t)$}   
  \put(-16,195){${\bf{d}}(t,\eta)$}   
  \put(-7,173){$\eta$}   
  \put(-6,152){${\bf{p}}(t)$}   
  \put(-80,170){${\bf{t}}(t)$}   
  \put(-6,52){$\theta$}   
  \put(-75,95){${\bf{n}}(s)$}   
  \put(-13,82){${\bf{d}}(s,\theta)$}   
  \put(-11,14){${\bf{p}}(s)$}   
  \put(-79,57){${\bf{t}}(s)$}   
\caption{(a) The schematic of a section of the edge of an open bilayer, (b) cross-sections of the edge at positions ${{\bf{x}}}(s)$ and ${{\bf{x}}}(t)$ with Darboux frame $\{{\bf{t}}, {\bf{n}}, {\bf{p}}\}$ at those positions.}
\label{Fig2}
\end{figure}

On the boundary curve $\cal C$ of the surface $\cal S$, the unit normal to the surface is denoted by $\bfn$. Also, since $\dot\bfx$ is a unimodular vector, its arclength derivative $\ddot\bfx$ is perpendicular to $\dot\bfx$, and thus, can be considered as a linear combination
\begin{equation}
\ddot\bfx=\kn\bfn+\kg\,\bfn\times\dot\bfx,
\label{accel}
\end{equation}
of the unit normal $\bfn$, and the product $\bfn\times\dot\bfx$. Notice that the unit normal $\bfn$ to the surface ${\cal S}$ is different from the unit normal $\bf{N}$ of the curve ${\cal C}$. Further, let $\bfp$ denote the unit vector in the tangent plane of ${\cal S}$ perpendicular to the unit tangent $\bft$ while pointing outward. We call $\bfp$ the unit tangent-normal. The set of unit normal $\bfn$ to the surface ${\cal S}$ at ${\cal C}$, unit tangent-normal $\bfp$, and the unit tangent $\bft$, which is $\bft=\bfn\times\bfp$, form the oriented basis $\{\bft,\bfn,\bfp\}$ on ${\cal C}$, known as the Darboux frame (Figure~\ref{Fig1}). Considering that the normal $\bf{N}$ and binormal $\bf{B}$ of the Frenet frame of $\cal C$ are also perpendicular to $\bft$, they both lie in the plane spanned by the normal $\bfn$ and tangent-normal $\bfp$ of the Darboux frame. Therefore, they are related to $\bfn$ and $\bfp$ by 
\begin{equation}
\left.
\begin{split}
\bf{N}&=(\cos\psi)\bfn-(\sin\psi)\bfp,
\\[4pt]
\bf{B}&=(\sin\psi)\bfn+(\cos\psi)\bfp,
\end{split}
\,\right\}
\label{decompose}
\end{equation}
where $\psi$ denotes the angle between the unit normal $\bf{N}$ of the curve $\cal C$ and the unit normal $\bfn$ to the surface $\cal S$. Following the proof provided in~\ref{appproof}, derivatives of $\bft$, $\bfn$, and $\bfp$ with respect to the arclength $s$ along ${\cal C}$ are expressed as
\begin{equation}
\left[ \begin{array}{c}{\dot{\bft}} \\{\dot{\bfn}} \\ {\dot\bfp} \end{array} \right]=\begin{bmatrix}0 & \kappa\cos\psi & -\kappa\sin\psi \\-\kappa\cos\psi & 0 & \tau+\dot\psi \\\kappa\sin\psi & -\tau-\dot\psi & 0 \\\end{bmatrix} \left[\begin{array}{c} {\bft} \\{\bfn} \\ {\bfp} \end{array} \right].
\label{frenetserret}
\end{equation}
The quantity 
\begin{equation}
\label{geodestor}
\tau_g=\tau+\dot\psi,
\end{equation}
 is called \emph{geodesic torsion} of the curve $\cal C$ on $\cal S$. This quantity describes the rate of the rotation of the tangent plane of the surface $\cal S$ about the unit tangent to the curve $\cal C$ with respect to the arc length $s$~\cite{guggenheimer1977differential}. Also, $\tau_g$ can be expressed alternatively as
\begin{equation}
\tau_g=\dot\bfx\cdot(\bfn\times\dot\bfn).
\end{equation}
Further, the curvature vector $\bfk$ of the curve $\cal C$ on the surface $\cal S$ is the sum of the normal curvature vector $\bfkn$, and the tangential (or geodesic) curvature vector $\bfkg$, i.e.
\begin{equation}
\bfk=\bfkn+\bfkg.
\label{curvvec}
\end{equation}
The normal curvature vector $\bfkn$ is the projection of the curvature vector $\bfk$ along the normal $\bfn$ of the surface ${\cal S}$. The geodesic curvature $\bfkg$ is perpendicular to the unit normal $\bfn$ to the surface, and, thus, lies in the tangent plane of the surface ${\cal S}$. Hence,
\begin{equation}
\bfkn=(\bfk\cdot\bfn)\,\bfn,\quad\text{and}\quad \bfkg=\bfn\times(\bfk\times\bfn).
\end{equation}
According to~\eqref{b2}, the curvature vector $\bfk$ of $\cal C$ is 
$\bfk=\kappa\bf{N}$. Since $\psi$ denotes the angle between $\bf{N}$ and $\bfn$, the magnitude $\kappa_n$ of the normal curvature $\bfkn$ is 
\begin{equation}
\label{accelnew}
\kappa_n=\bfk\cdot\bfn=\kappa\,{\bf{N}}\cdot\bfn=\kappa\cos\psi.
\end{equation}
The magnitude $\kg$ of the tangential (or geodesic) curvature vector $\bfkg$, is a bending invariant, and is given by 
\begin{equation}
\label{kgfinal}
\kg=-\bfp\cdot\dot{\bft}=-\kappa\,\bfp\cdot{\bf{N}}=({\bfn}\times{\bft})\cdot\dot{\bft}.
\end{equation}
According to $\dot\bft=\kappa\bf{N}$, and 
\begin{equation}
\bfn=(\cos\psi)\bf{N}+(\sin\psi)\bf{B},
\label{framede2}
\end{equation}  
the right-hand side of~\eqref{kgfinal} results
\begin{equation}
\bfkg=\kg\,(\bfn\times\bft),\quad\text{where}\quad\kg=\kappa\sin\psi.
\label{curvvec22}
\end{equation}  
As mentioned earlier, the geodesic curvature vector $\bfkg$ at any point of the curve ${\cal C}$ on the surface ${\cal S}$ is the vectorial projection of the curvature vector $\bfk$ of the curve ${\cal C}$ into the tangent plane of the surface ${\cal S}$ at that point. This quantity is an intrinsic property of the surface, which reflects the deviation of the curve ${\cal C}$ from a geodesic on the surface ${\cal S}$~\cite{synge1969tensor}. In general, for a geodesic, the geodesic curvature $\kappa_g$ at any point is zero. Further, for a geodesic, the unit normal $\bf{N}$ of the curve ${\cal C}$ coincides with the unit normal $\bfn$ of the surface ${\cal S}$, or, equivalently, the osculating plane of ${\cal C}$ at each point is perpendicular to the tangent plane of the surface ${\cal S}$ at that point~\cite{guggenheimer1977differential}. This means that the Darboux frame and the Frenet frame for a geodesic are the same at any point.

According to~\eqref{geodestor}, and the right-hand sides of~\eqref{accelnew} and~\eqref{curvvec22}, the arclength derivatives of $\{\bft,\bfn,\bfp\}$ in~\eqref{frenetserret} take the form 
\begin{equation}
\left.
\begin{split}
\dot\bft&=\kappa_n\bfn-\kappa_g\bfp,
\\
\dot\bfn&=-\kappa_n\bft+\tau_g\bfp,
\\
\dot\bfp&=\kappa_g\bft-\tau_g\bfn.
\label{framede}
\end{split}
\right\}
\end{equation}  

\section{Modeling assumptions}\label{assumptions}

 The phospholipid molecules comprising the edge are allocated so that their hydrophilic parts lie on a thin semicylindrical surface as shown in Figure~\ref{Fig2} to form a core shielding the hydrophobic tails from the surrounding solution. The centerline of the edge is denoted by a boundary curve $\cal{C}$. The following assumptions, which are based on the previously reported observations~\cite{jiang2004molecular,wohlert2006free,karatekin2003cascades,guven2014terasaki,zhelev1993tension}, are considered to model the edge of an open lipid bilayer:
\begin{enumerate}[(i)\,$\cdot$]
\item The phospholipid molecules comprising the edge are modeled as one-dimensional rigid rods of the same length $a$.
\item The lipid molecules are assumed to be perpendicular to the centerline $\cal{C}$, residing in the plane spanned by the unit normal $\bf{n}$ and the unit tangent-normal $\bf{p}$, as depicted in Figure~\ref{Fig2}. This assumption is valid as long as the concentration of the lipid molecules on the edge $\cal C$ is sufficiently high.
\item The phospholipid molecules at any cross-section of the edge have uniform angular distribution. 
\item The distribution of the phospholipid molecules at any point along $\cal{C}$ is denoted by the molecular density function $\Pi>0$. In contrast to the angular distribution,
which is assumed to be uniform because of symmetry considerations, the molecular distribution along $\cal C$ may be nonuniform as a result of localized curvature.
\end{enumerate}

Consider a lipid molecule at the position corresponding to $s$ on $\cal{C}$ with orientation $\theta$ measured counterclockwise from the corresponding tangent-normal ${\bfp(s)}$, as depicted schematically in Figure~\ref{Fig2}b. Let the director $\bfd(s,\theta)$ denote the orientation of this molecule. By the second assumption, such a director can be expressed as a linear combination 
\begin{equation}
\bfd(s,\theta)=(\cos\theta)\,\bfp(s)+(\sin\theta)\,\bfn(s),
\label{dir1}
\end{equation}
where $\bfp(s)$ and $\bfn(s)$ denote the tangent-normal and the unit normal (to ${\cal S}$) at the position corresponding to $s$.

\section{Derivation of the free-energy density}\label{derivation}

In this section, the free-energy density of the edge of an open lipid bilayer is derived taking into account the interactions between the molecules comprising the edge. To do so, a microphysical approach is applied, guided by the work of Keller and Merchant~\cite{keller1991flexural}. 

Consider two molecules, with directors $\bfd$ and $\bfd'$, located respectively at positions $\bfx$ and $\bfx'$ interior to ${\cal C}$. Let the interaction energy (encompassing steric, electrostatic, and other relevant effects) between the molecules under consideration be denoted by
\begin{equation}
\label{2121}
\Omega(\bfx,\bfx',\bfd,\bfd').
\end{equation}
Following Keller and Merchant~\cite{keller1991flexural}, we assume that the interaction energy between two molecules separated by more than a fixed cutoff distance $\delta$ vanishes, in which case
\begin{equation}
\label{condition}
\Omega(\bfx,\bfx',\bfd,\bfd')=0
\quad\text{if}\quad\abs{\bfx-\bfx'}>\delta. 
\end{equation}
In the present setting, the cutoff distance $\delta$ is required to be small relative to the characteristic length $\ell$ of the edge, so that a dimensionless measure $\varrho$ of cutoff distance obeys
\begin{equation}
\varrho:=\frac{\delta}{\ell}\ll1.
\label{epsdef}
\end{equation}
Hereafter, we restrict attention to interaction energies $\Omega$ that are of the form~\eqref{2121} but are also frame indifferent~\cite{truesdell2004non}. It then follows that $\Omega(\bfx,\bfx',\bfd, \bfd')$ may depend on the positions $\bfx$ and $\bfx'$ and the directors $\bfd$ and $\bfd'$ only through the length $\abs{\bfx-\bfx'}$ of the vector between $\bfx$ and $\bfx'$, the dot products $(\bfx-\bfx')\cdot\bfd$ and $(\bfx-\bfx')\cdot\bfd'$ formed by the directors and that vector, and the dot product $\bfd\cdot\bfd'$ formed by the directors. Like Keller and Merchant~\cite{keller1991flexural}, we assume that dependence of the interaction energy on the length of the relative position vector is scaled by the ratio $\varrho$ defined in~\eqref{epsdef} and, thus, that
\begin{equation}
\Omega(\bfx,\bfx',\bfd,\bfd')
=2\,\tilde\Omega(\varrho^{-2}\bfr\cdot\bfr,\bfr\cdot\bfd,\bfr\cdot\bfd',\bfd\cdot\bfd'),
\label{212}
\end{equation}
with $\bfr=\bfx-\bfx'$. The factor of two in the right-hand side of~\eqref{212} is for simplifying later calculations. Notice that $\Omega$ depends explicitly on $\delta$, whereas $\tilde\Omega$ does not. Consequently,~\eqref{condition},~\eqref{epsdef}, and~\eqref{212} yield
\begin{equation}
\label{ieradius}
\tilde\Omega(s^2,\rho_1,\rho_2,\rho_3)=0
\quad\text{if}\quad
s>\ell,
\end{equation}
where
\begin{equation}
\rho_1=s\varrho\,\hat{\bfr}\cdot\bfd,
\quad
\rho_2=s\varrho\,\hat{\bfr}\cdot\bfd',
\quad
\rho_3=s\varrho\,\bfd\cdot\bfd',
\label{arguments}
\end{equation}  
with $\hat\bfr$ being the unit vector corresponding to the intermolecular vector $\bfr$.

As a consequence of the foregoing discussion, the net free-energy $\phi_{\text{net}}$ of the edge can be expressed as
\begin{equation}
\phi_{\text{net}}=\int_{0}^{L}\frac{1}{2}\bigg(\int_{-\frac{\pi}{2}}^{\frac{\pi}{2}}\mskip-3mu\omega(s,\theta)\dtheta\bigg)\Pi(s)\ds,
\label{energy0}
\end{equation}
where
\begin{equation}
\omega(s,\theta)
=\int_{0}^{L}\mskip-11mu\int_{\frac{-\pi}{2}}^{\frac{\pi}{2}}
\Omega\big(\bfx(s),\bfx(t),\bfd(s,\theta),\bfd(t,\eta)\big)\Pi(t)\deta\dt
\label{interaction0}
\end{equation}
is the free energy due to the interactions between the molecule with director $\bfd(s,\theta)$ at $\bfx(s)$ with all other molecules and where a factor of one-half compensates for the double counting of interactions arising from integrating over both $s$ and $t$ from $0$ to $L$. From~\eqref{energy0}, the free-energy density $\phi$ at position $\bfx(s)$ on $\cal{C}$ is simply
\begin{equation}
\phi=\frac{1}{2}\bigg(\int_{\frac{-\pi}{2}}^{\frac{\pi}{2}}\mskip-3mu\omega(s,\theta)\dtheta\bigg)\Pi(s).
\label{e11}
\end{equation}
The function $\Pi$ denotes the density of the lipid molecules at any point of the curve $\cal C$. Following the proof which relies on the Taylor series expansion of the integrand of~\eqref{e11} with respect to $\varrho$ up to the second derivative term, provided in~\ref{Derivation of the free-energy density},~\eqref{e11} becomes
\begin{equation}
\phi=k_\circ+k_1\kg^2+k_2\kn^2+k_3\kg+k_4\kn \\
+k_5\kn\kg+k_6\tau_g^2,
\label{finalform}
\end{equation}
which includes a quadratic expression in terms of $\kappa_g$, $\kappa_n$, and $\tau_g$. Also, $k_\circ$ and the coefficients $k_i$ in~\eqref{finalform} are provided in~\ref{Derivation of the free-energy density}. Notice that $k_\circ$ is the standard line energy of the edge---the part which is independent of edge geometry---while the coefficients $k_i$, $i=1,2,5,6$ represent the flexural and torsional rigidities of the edge.
Up to the second derivative term of the Taylor expansion considered here, the derived model~\eqref{finalform} contains the linear terms of the normal curvature $\kn$ and the geodesic curvature $\kg$ of the boundary curve $\cal{C}$, while it does not incorporate the linear term of the geodesic torsion $\tau_g$ of $\cal{C}$. Considering a simplification of~\eqref{finalform} in the form
\begin{equation}
\phi={k_\circ}+{k_1}(\kg-{\kappa_{g\circ}})^2+{k_2}(\kn-{\kappa}_{n\circ})^2\\
+{k_5}{\kn\kg}+{k_6}{\tau_g}^2,
\label{finalform3}
\end{equation}
where
\begin{equation}
\begin{split}
{\kappa_{g\circ}}=-\frac{k_3}{2k_1},
\quad
{\kappa}_{n\circ}=-\frac{k_4}{2k_2},
\end{split}
\label{dir111}
\end{equation}
it can be inferred that the remaining coefficients $k_3$ and $k_4$ are related to the spontaneous geodesic and normal curvatures $\kappa_{g\circ}$ and $\kappa_{n\circ}$ of ${\cal{C}}$. This transpires that our model captures the spontaneous normal and geodesic curvatures ${\kappa}_{n\circ}$ and ${\kappa}_{g\circ}$ of the edge, while an spontaneous geodesic torsion is absent from the free energy. Further,~\eqref{finalform} and ~\eqref{finalform3} include the coupling of the normal curvature $\kn$ and the geodesic curvature $\kg$ via the term $\kn\kg$. However, the couplings of the geodesic torsion $\tau_g$ with the normal and geodesic curvatures $\kn$ and $\kg$ are absent. Our calculations show that the latter terms only appear by including higher order terms in the Taylor expansion, and hence, are of less significance. In addition, if the molecules have non-uniform angular distribution, i.e. the molecular distribution function $\Pi$ is allowed to depend upon $\theta$ or $\eta$, the model would also include a linear term in geodesic torsion $\tau_g$ that would lead to the presence of a spontaneous geodesic torsion.

The first term $k_{\circ}$ on the right-hand side of~\eqref{finalform3} is insensitive to the shape of the boundary. Since the molecular distribution on the boundary has implicit dependence upon the ambient temperature and concentration, these effects may be encompassed in $k_{\circ}$ and in the moduli $k_1$--$k_6$. The line tension $k_{\circ}$ has been obtained through experiments and molecular dynamic simulations for various types of lipid bilayers (see~\cite{jiang2004molecular} for a comparison of different measurements). However, there is not enough literature on measurement of the remaining coefficients. By fitting our model to existing measurements of the line tension, the controlling parameters of the interaction potential can be evaluated, and further used to obtain the remaining coefficients in~\eqref{finalform}.

The derived model~\eqref{finalform} can be simplified into the previously presented models for the free-energy of the edge. In particular, the general form~\eqref{finalform} provides a development to the theoretical investigations of open lipid bilayers presented by Tu and Ou-Yang~\cite{tu2004geometric}, and Guven {\etal}~\cite{guven2014terasaki}.

\subsection{Total free-energy of the edge}

The net free-energy $\phi_{\text{net}}$ associated with the elasticity of the edge of the open lipid bilayer is simply obtained by integrating the free-energy density $\phi$ in~\eqref{finalform3} over the centerline ${\cal C}$ of the edge by
\begin{equation}
\phi_{\text{net}}=\int_{0}^{L}\phi\ds.
\label{finaltot}
\end{equation}

\section{Applying a concrete interaction potential}\label{results}

The interaction potential in the model developed in the previous section was assumed to be a general function of four frame-indifferent arguments in terms of the intermolecular vector and the orientation of phospholipid molecules. There are numerous concrete models for such interaction potentials between axisymmetric particles, which are vastly employed for numerical simulations of liquid crystals and other similar systems. Our derivation gives rise to integral representations for elastic moduli $k_\circ$--$k_6$ of the edge. Substituting the general form $\tilde{\Omega}$ in~\eqref{dir3} with an available interaction potential yields the material parameters $k_\circ$--$k_6$ appearing in~\eqref{finalform}.

\begin{figure} [t]
 \centering
 \includegraphics [height=1in] {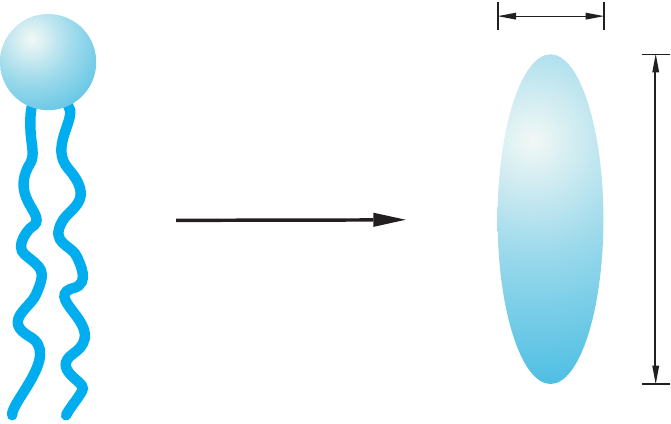}
 \put(2,34){$\sigma_e$}   
 \put(-24,75){$\sigma_s$}   
\caption{(a) The schematic of a phospholipid molecule modelled as an ellipsoidal particle.} 
\label{Fig3}
\end{figure}

One of the standard examples among such pair interaction potentials is the spheroidal-particle model proposed by Berne and Pechukas~\cite{berne1972gaussian}, and Gay and Berne~\cite{gay1981modification}, in which the molecules are approximated by ellipsoids of revolution, or spheroids (see Figure~\ref{Fig3}). According to such model, the interaction potential between two molecules with the intermolecular vector $\bfr$ and the directors $\bfd$ and $\bfe$ possesses the multiplicative decomposition
\begin{equation}
\label{Berne1}
\tilde\Omega(\bfr,\bfd,\bfe)=\xi(\hat\bfr,\bfd,\bfe)\zeta(\bfr,\bfd,\bfe),
\end{equation}
where $\xi(\hat\bfr,\bfd,\bfe)$ and $\zeta(\bfr,\bfd,\bfe)$ denote the strength and distance parameters respectively. The strength parameter $\xi(\hat\bfr,\bfd,\bfe)$ depends upon the orientation of the molecules and that of the intermolecular vector $\bfr$ through (Gay and Berne~\cite{gay1981modification})
\begin{multline}
\label{spn}
\xi(\hat\bfr,\bfd,\bfe)=\frac{4\,\xi_{\circ}}{(1-\chi^2(\bfd\cdot\bfe)^2)^{\nu/2}}\\
\Big[1-\frac{\chi'}{2}\Big(\frac{(\hat\bfr\cdot\bfd+\hat\bfr\cdot\bfe)^2}{1+\chi'\bfd\cdot\bfe}+\frac{(\hat\bfr\cdot\bfd-\hat\bfr\cdot\bfe)^2}{1-\chi'\bfd\cdot\bfe}\Big)\Big]^\mu,
\end{multline}
with $\nu$, $\mu$, and $\xi_{\circ}$ the fitting parameters to be chosen. More specifically, $\nu$ depends upon the arrangement type of the molecules (e.g. side-to-side or end-to-end), whereas $\xi_\circ$ is a constant that specifies the kind of molecules under consideration. The parameter $\chi$ in~\eqref{spn} is the shape anisotropy parameter, given in terms of the ratio of the length $\sigma_e$ to the breadth $\sigma_s$ of the molecules by
\begin{equation}
\label{GBC}\chi=\frac{(\sigma_e/\sigma_s)^2-1}{(\sigma_e/\sigma_s)^2+1}.
\end{equation}
Also, the parameter $\chi'$ in~\eqref{spn} is given by
\begin{equation}
\label{chi} 
\chi'=\frac{(\eps_e/\eps_s)^{1/\mu}-1}{(\eps_e/\eps_s)^{1/\mu}+1},
\end{equation}
where $\eps_e$ and $\eps_s$ denote the strength parameters for end-to-end and side-to-side arrangement of the molecules respectively. The distance parameter $\zeta(\bfr,\bfd,\bfe)$ in~\eqref{Berne1} is given by (Berne and Pechukas~\cite{berne1972gaussian})
\begin{equation}
\label{sp4}
\zeta(\bfr,\bfd,\bfe)=\exp\bigg({\frac{-\abs{\bfr}^2}{\varsigma^2(\hat\bfr,\bfd,\bfe)}}\bigg),
\end{equation}
where $\varsigma(\hat\bfr,\bfd,\bfe)$ is called the range parameter and is given as a function of the orientation of the molecules and that of the intermolecular vector $\bf{r}$ by
\begin{equation}
\label{sp2}
\varsigma(\hat\bfr,\bfd,\bfe)=\sigma_{\circ}\Big[1-\frac{\chi}{2}\Big(\frac{(\hat\bfr\cdot\bfd+\hat\bfr\cdot\bfe)^2}{1+\chi\bfd\cdot\bfe}+\frac{(\hat\bfr\cdot\bfd-\hat\bfr\cdot\bfe)^2}{1-\chi\bfd\cdot\bfe}\Big)\Big]^\frac{-1}{2}.
\end{equation}
In~\eqref{sp2}, $\sigma_{\circ}$ is related to the breadth of the molecules, $\sigma_s$, via $\sigma_{\circ}=\sqrt{2}\sigma_s$. Following Whitehead {\etal}~\cite{whitehead2001molecular}, the parameters $\mu$ and $\nu$ are chosen as 
\begin{equation}
\nu=-1,\quad\text{and}\quad \mu=2.\qquad 
\end{equation}

By applying the interaction potential~\eqref{Berne1} in~\eqref{proof}, and assuming a constant molecular density $\Pi$ along the boundary $\cal C$, the coefficients $k_i$ in~\eqref{finalform} are obtained as
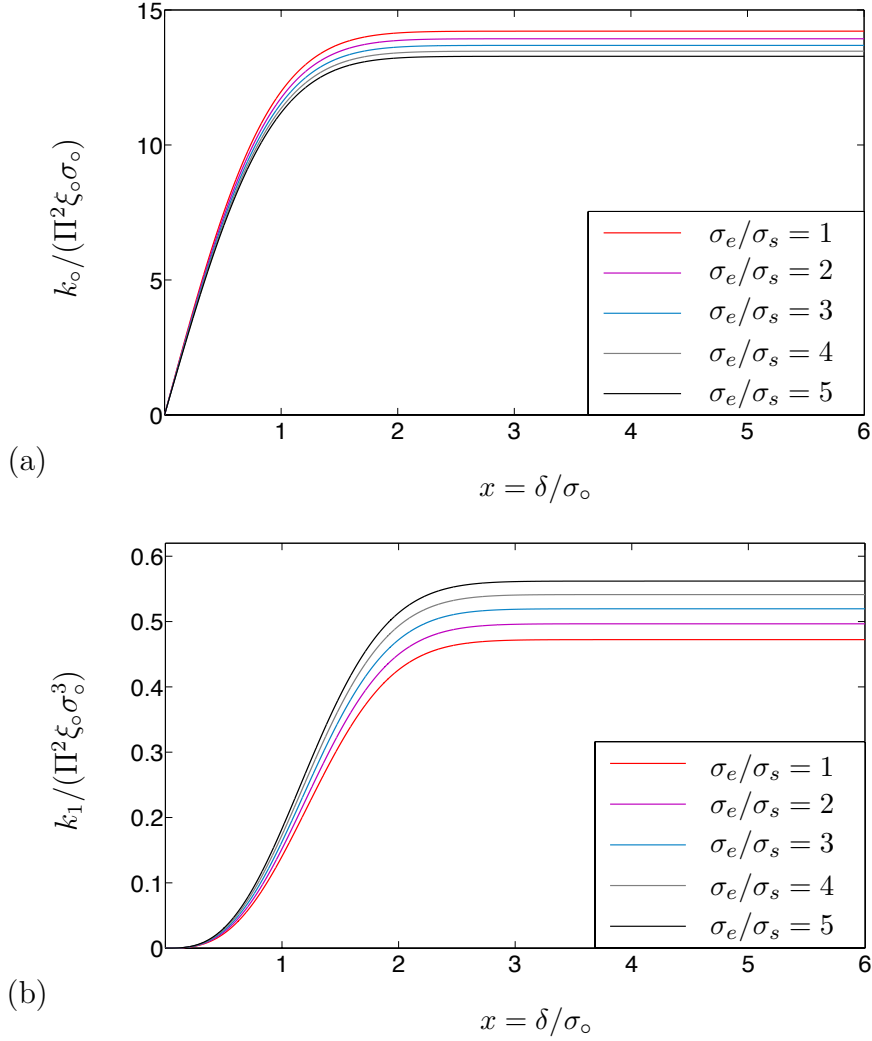
\begin{figure}
\begin{center}
\subfigure{(a)}{\resizebox*{12cm}{!}{\includegraphics{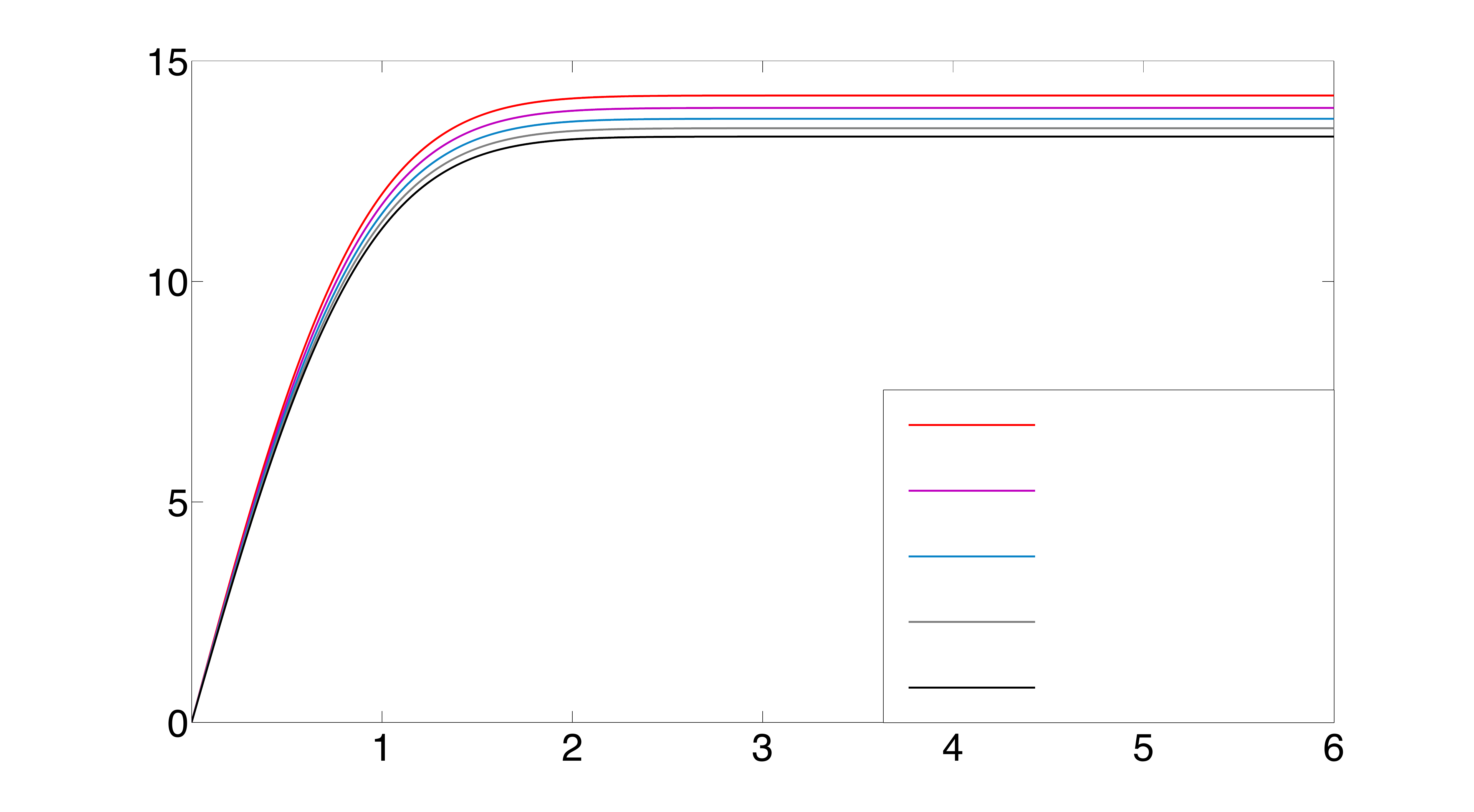}}}
\put(-340,65){${\begin{sideways}{$k_{\circ}/(\Pi^2\xi_{\circ}\sigma_{\circ})$}\end{sideways}}$} 
\put(-178,-10){\small$x=\delta/\sigma_{\circ}$}
\put(-91,26){\small$\sigma_e/\sigma_s=5$}
  \put(-91,41){\small$\sigma_e/\sigma_s=4$}
  \put(-91,57){\small$\sigma_e/\sigma_s=3$}
  \put(-91,72){\small$\sigma_e/\sigma_s=2$}
  \put(-91,86){\small$\sigma_e/\sigma_s=1$}\\
\subfigure{(b)}{\resizebox*{12cm}{!}{\includegraphics{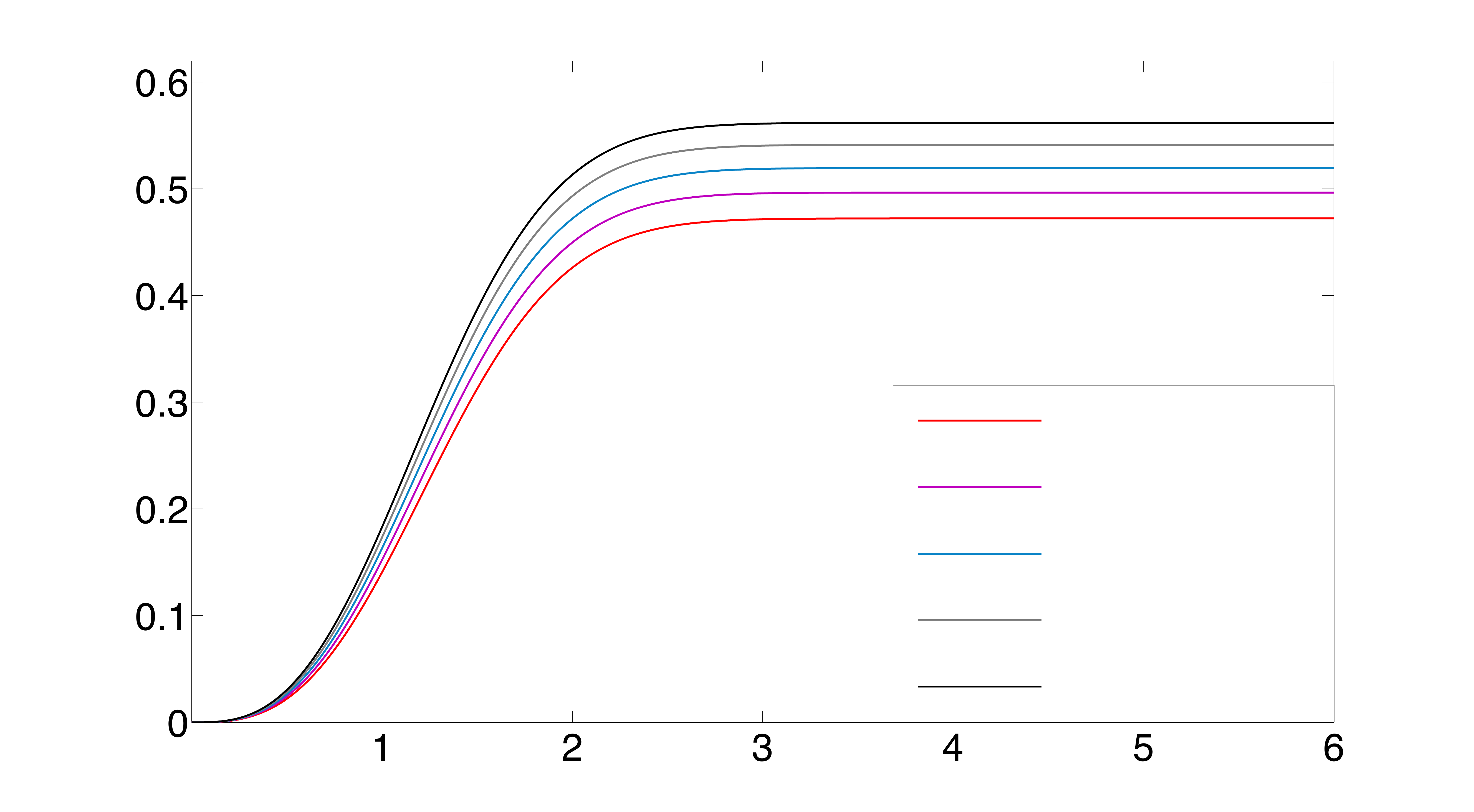}}}
 \put(-340,65){\begin{sideways}{${k_1}/({\Pi^2{\xi_{\circ}}{\sigma^{3}_{\circ}})}$}\end{sideways}}
\put(-178,-10){\small$x=\delta/\sigma_{\circ}$}
 \put(-91,26){\small$\sigma_e/\sigma_s=5$}
  \put(-91,41){\small$\sigma_e/\sigma_s=4$}
  \put(-91,57){\small$\sigma_e/\sigma_s=3$}
  \put(-91,72){\small$\sigma_e/\sigma_s=2$}
  \put(-91,86){\small$\sigma_e/\sigma_s=1$}
\caption{(a). Schematic of the constant line energy ${{k_{\circ}}/{\Pi^2\xi_{\circ}\sigma_{\circ}}}$ in terms of the dimensionless cut-off distance $\delta/\sigma_{\circ}$. (b) Schematic of the flexural rigidity ${{k_1}/{\Pi^2\xi_{\circ}\sigma^3_{\circ}}}$ in terms of the dimensionless cut-off distance $\delta/\sigma_{\circ}$. As is evident from the plots, the change in ${k_{\circ}}/{\Pi^2\xi_{\circ}\sigma_{\circ}}$ and ${{k_1}/{\Pi^2\xi_{\circ}\sigma^3_{\circ}}}$ is negligible after $\delta=3\sigma_{\circ}$. Consequently, the effective cut-off distance after which the potential decays rapidly, can be reasonably approximated by $\delta=3\sigma_{\circ}$.}
\label{Fig4}
\end{center}
\end{figure}
\begin{align}
&\notag\frac{k_\circ}{\Pi^2\xi_{\circ}\sigma_{\circ}}=\sqrt{\pi} \erf(x)I_\circ,
\\[4pt]
&\notag\frac{k_1}{\Pi^2\xi_{\circ}\sigma^3_{\circ}}=\frac{k_2}{\Pi^2\xi_{\circ}\sigma^3_{\circ}}=\frac{\sqrt{\pi}}{128}\erf(x)\Big(\chi^2J+(\chi^2+2)I\Big)\\
&\notag\qquad+\frac{xe^{-x^2}}{192}\Big(\chi^2(J+I)(2x^2-3)-2I(2x^2+3)\Big),\\[4pt]
& k_3=k_4=k_5=k_6=0.
\label{coefficients1}
\end{align}  
where $x$ is a dimensionless parameter defined as the ratio of the cut-off distance $\delta$ to $\sigma_{\circ}$ as
\begin{equation}
x=\frac{\delta}{\sigma_{\circ}},
\end{equation}
and $I_\circ$, $I$ and $J$ are integral representations shown in~\ref{Appendix2}. Hence, the free-energy density $\phi$ in~\eqref{finalform} specializes to
\begin{equation}\label{nresult}
\phi=k_\circ+k_1(\kappa^2_n+\kappa^2_n)=k_\circ+k_1\,\kappa^2.
\end{equation}

\begin{figure}
\begin{center}
\subfigure{(a)}{\resizebox*{12cm}{!}{\includegraphics{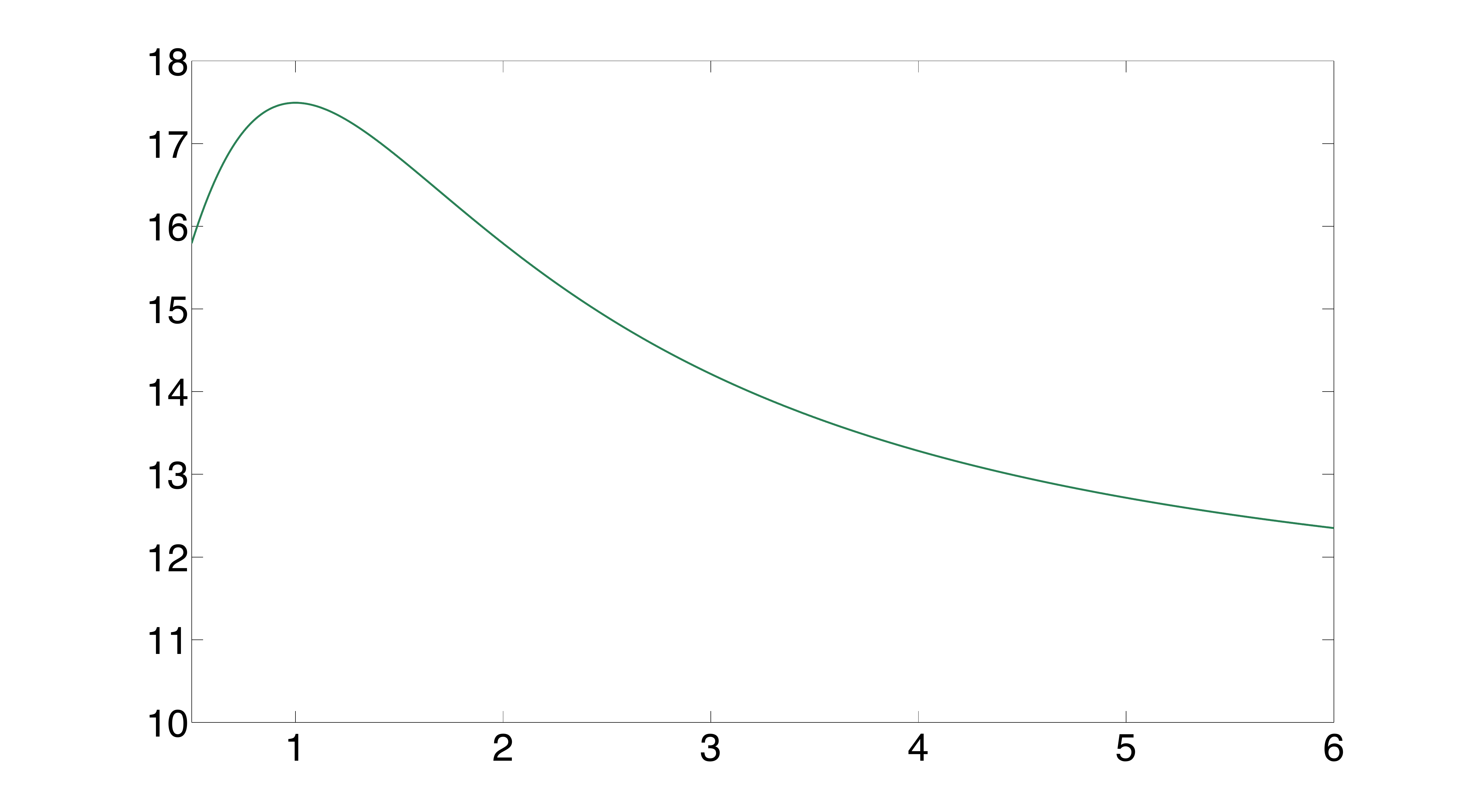}}}
\put(-178,-13){$\sigma_e/\sigma_{s}$}   
 \put(-350,85){\begin{sideways}\large$\frac{k_{\circ}}{\Pi^2\xi_{\circ}\sigma_{\circ}}$\end{sideways}}  \\ 
\subfigure{(b)}{\resizebox*{12cm}{!}{\includegraphics{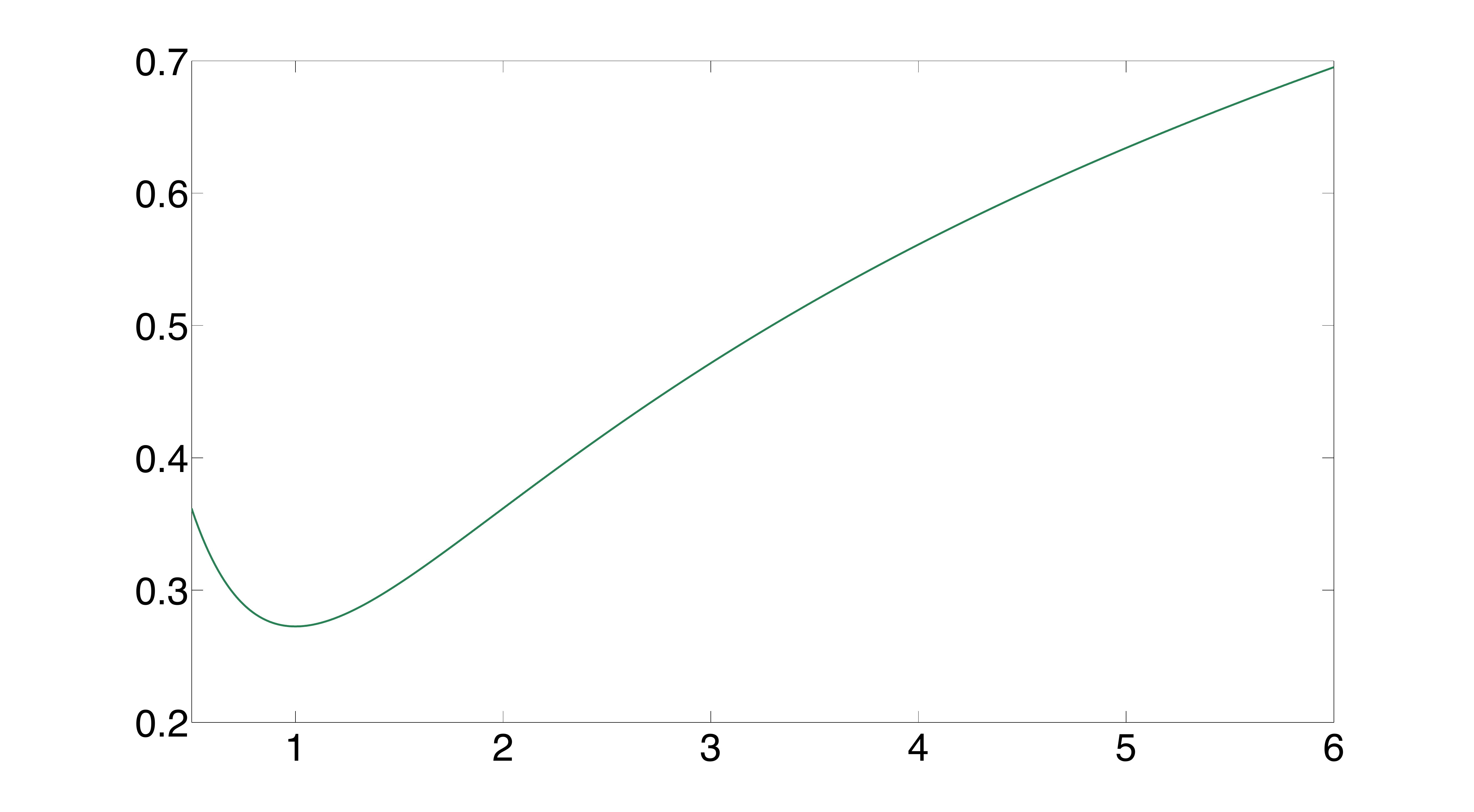}}}
   \put(-178,-13){$\sigma_e/\sigma_{s}$}   
 \put(-350,55){\begin{sideways}\large$\frac{k_1}{\Pi^2\xi_{\circ}\sigma^3_{\circ}}=\frac{k_2}{\Pi^2\xi_{\circ}\sigma^3_{\circ}}$\end{sideways}}
\caption{(a). Schematic of the constant line energy $k_{\circ}/{\Pi^2\xi_{\circ}\sigma_{\circ}}$ in terms of the aspect ratio $\sigma_e/\sigma_{s}$ for a cut-off distance $\delta/\sigma_{\circ}=3$. (b) Schematic of the flexural rigidity $k_1/\Pi^2\xi_{\circ}\sigma_{\circ}$ in terms of the aspect ratio $\sigma_e/\sigma_{s}$ for the cut-off distance $\delta=3\sigma_{\circ}$.}
\label{Fig5}
\end{center}
\end{figure}

A single phospholipid molecule can be envisioned as a molecule in which a water-soluble spherical head is attached to a pair of water-insoluble tails. Here we rely on the dimensions of a specific kind of phospholipid molecule (DPPC/Water system) reported by Mashaghi {\etal}~\cite{mashaghi2012hydration} According to their estimation, the length of the aforementioned phospholipid molecule from the center of the head-group to the tail is $\sim$ 22.5--30 $A^\circ$, and the diameter of the head-group is $\sim$ 7--10 $A^\circ$. The total volume of that molecule is thus the sum of the volume of the spherical head and that of a cylindrical tail-group. Based on the equality of the volumes of the phospholipid molecule and that of the spheroidal replacement, the aspect ratio $\sigma_e/\sigma_s$ of the spheroid in Figure~\ref{Fig3} is obtained between 3 and 4. Hence, the schematic of the constant line energy $k_\circ$ and that of the flexural rigidity $k_1$ are depicted in terms of the dimensionless cutoff distance $\delta/\sigma_\circ$ in Figure~\ref{Fig4}, for molecular aspect ratio $\sigma_e/\sigma_s$ between 1 and 5. According to Figure~\ref{Fig4}, the change in the constant line energy $k_\circ$ and flexural rigidity $k_1$ is negligible after some value of the dimensionless cutoff distance $\delta/\sigma_\circ$. Therefore, a rather conservative choice for the cut-off distance, which guarantees inclusion of all significant molecular interactions, is
\begin{equation}
{\delta^*}=3\sigma_{\circ}.
\label{cutof} 
\end{equation}
This result has been used to obtain the constant part of the line energy $k_{\circ}$ and flexural rigidity $k_1$ of the edge in terms of the molecular aspect ratio, as depicted in Figure~\ref{Fig5}.

\section{Illustrative example: dependence of free-energy on the pore size}\label{Illustrative}

\begin{figure}
  \centering
    \includegraphics [width=0.3\textwidth] {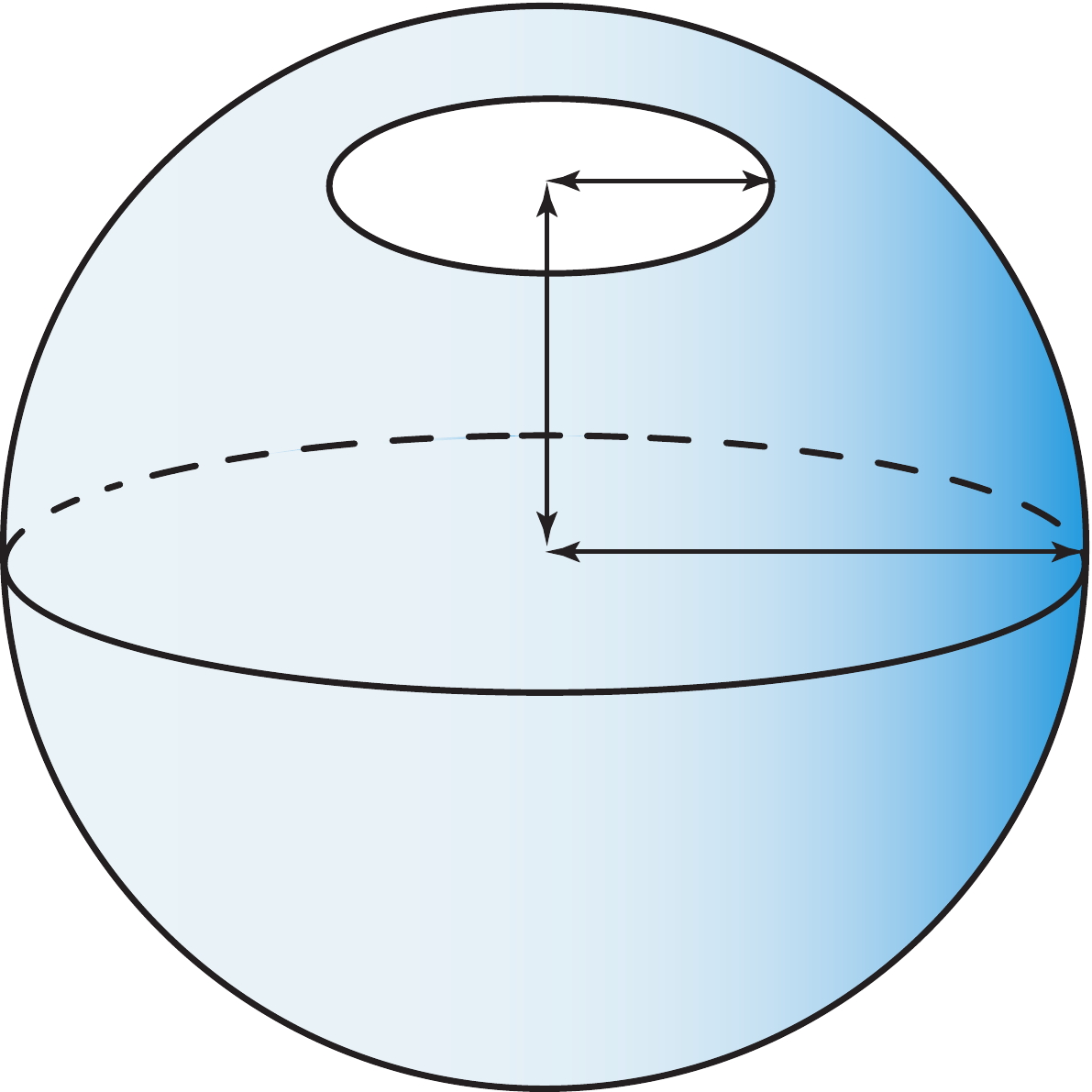}
 \put(-35,47){$R$} 
 \put(-47,91.5){$r$}
 \put(-67,75){$h$} 
\caption{Schematic of a pore on a spheroidal lipid bilayer} 
\label{Fig6}
\end{figure}

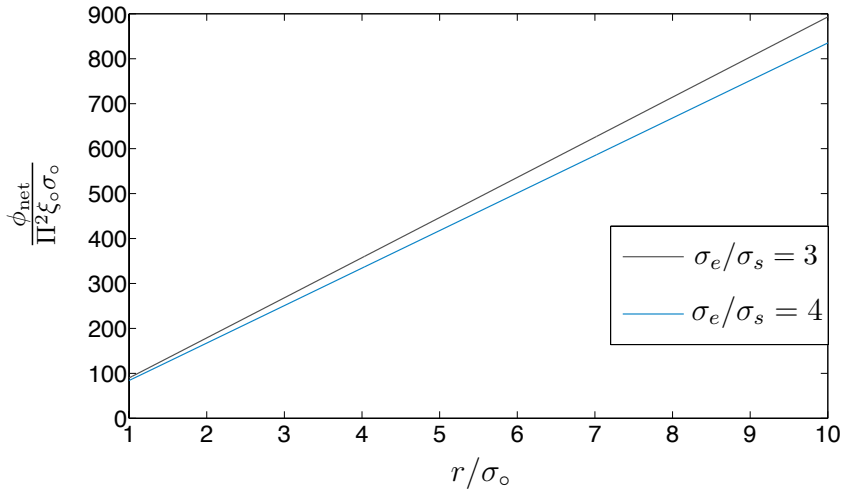
\begin{figure}
   \centering
  \includegraphics [height=2.6in] {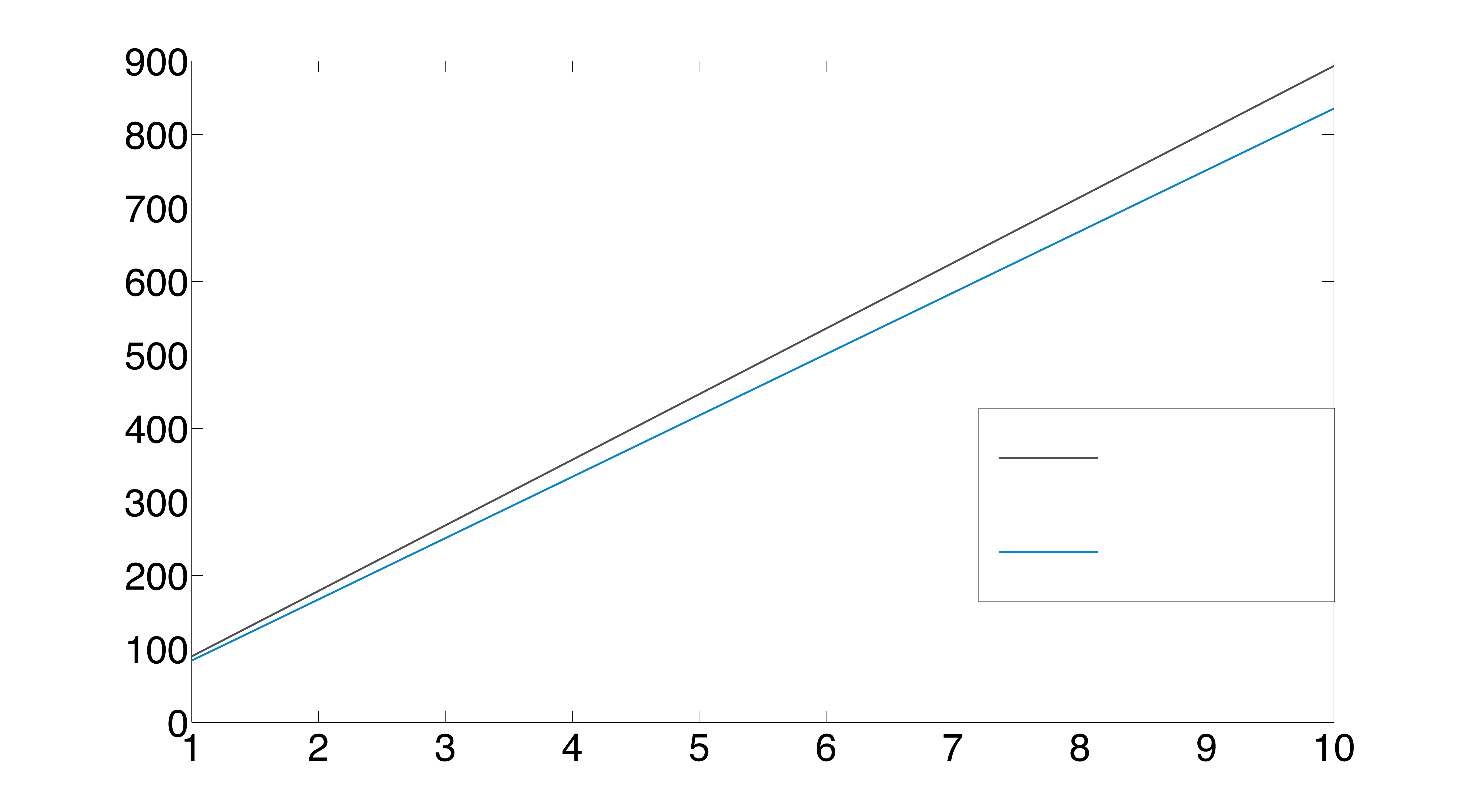}
  \put(-175,-3){$r/\sigma_{\circ}$}   
\put(-84,58){$\sigma_e/\sigma_s=4$}   
\put(-83,79){\small${\sigma_e}/{\sigma_s}=3$}   
\put(-343,85){\begin{sideways}\large$\frac{\phi_\text{net}}{\Pi^2\xi_{\circ}\sigma_{\circ}}$\end{sideways}} 
\caption{Schematic of the net free-energy ${{\phi_\text{net}}/{\Pi^2\xi_{\circ}\sigma_{\circ}}}$ versus the scaled pore-size $r/\sigma_\circ$ for two values of the aspect ratio $\sigma_e/\sigma_{s}$ and for the scaled cutoff distance $\delta=3\sigma_{\circ}$.} 
\label{Fig7}
\end{figure}

In order to estimate the change of energy of a pore with its size, consider the simple case of a spherical lipid bilayer with radius $R$, with a pore of radius $r$ at a distance $h$ from its centre, as depicted in Figure~\ref{Fig6}. For such a pore, the total curvature of the boundary curve $\cal C$ is $1/r$, and the geodesic torsion $\tau_g$ vanishes. Also, the normal and geodesic curvatures find the froms
\begin{equation}
\kn=\frac{1}{R}, \quad \text{and} \quad \kg=\sqrt{\kappa^2-\kn^2}=\frac{\sqrt{R^2-r^2}}{rR}.
\label{example2}
\end{equation}
As a result, the free-energy density~\eqref{finalform3} specializes to
\begin{equation}
\phi={k_\circ}+{k_1}\Big(\frac{\sqrt{R^2-r^2}}{rR}-{\kappa}_{g\circ}\Big)^2+{k_2}\Big(\frac{1}{R}-{\kappa}_{n\circ}\Big)^2\\
+{k_5}\Big(\frac{\sqrt{R^2-r^2}}{rR^2}\Big),
\label{example1}
\end{equation}
which with~\eqref{finaltot}, yields a representation for the net free-energy $\phi_{\text{net}}$ of the pore as
\begin{equation}
\frac{\phi_{\text{net}}}{2\pi r}={k_\circ}+{k_1}\Big(\frac{\sqrt{R^2-r^2}}{rR}-{\kappa}_{g\circ}\Big)^2+{k_2}\Big(\frac{1}{R}-{\kappa}_{n\circ}\Big)^2\\
+{k_5}\Big(\frac{\sqrt{R^2-r^2}}{rR^2}\Big).
\label{example2}
\end{equation}
It was demonstrated in the previous section that when the spheroidal interaction potential~\cite{berne1972gaussian} is employed in the microphysical model, the spontaneous curvatures ${\kappa}_{g\circ}$ and ${\kappa}_{n\circ}$, and the coefficient $k_5$ vanish, while the bending moduli $k_1$ and $k_2$ find the same value. Using those results in~\eqref{example2} yields
\begin{equation}
\frac{\phi_{\text{net}}}{2\pi}={k_\circ}r+\frac{k_1}{r},
\label{example3}
\end{equation}
whereby the energy of the pore does not depend on the size of the lipid bilayer, nor on the placing of the pore on it. For the cut-off distance $\delta^*=3\sigma_\circ$ and the molecular aspect ratios $\sigma_e/\sigma_s=3$ and $\sigma_e/\sigma_s=4$, the dependence of the net free-energy to pore size has been demonstrated in Figure~\ref{Fig7}. The second term in the right-hand side of~\eqref{example3} leads to a minimum point for the free energy for $r<\sigma_\circ$, which does not fall in the physically-relevant ranges of the pore size. For reasonable values of $r/\sigma_\circ$, the first term on the right of~\eqref{example3} is dominant and the dependence of the net energy on the pore size is effectively linear.

\section{Discussion and Summary}\label{discussion}

An expression for the free-energy density of the edge of an open lipid bilayer was derived taking into account the interaction between the constituent molecules. The resulting expression contains quadratic terms in geodesic curvature, normal curvature, and geodesic torsion of the boundary curve and a term including the multiplication of geodesic and normal curvatures.
The derived free-energy of the edge is the evidence of an excess energy due to the specific arrangement of the phospholipid molecules in the vicinity of the boundary of an open bilayer, in accord with the results of the existing molecular dynamic simulations~\cite{jiang2004molecular,wohlert2006free,de2006coarse,karatekin2003cascades}. Further, our study supplements the previous molecular dynamic simulations~\cite{jiang2004molecular,wohlert2006free,de2006coarse,karatekin2003cascades} and theories~\cite{may2000molecular} in which the free energy was obtained as a constant, by providing the contribution due to bending and torsional energies. For certain classes of lipid bilayers, the bending free-energy, which can be captured by our framework, is of more importance in contrast to the line energy~\cite{jiang2007simulations,pera2015edge}. In addition, our microphysical model justifies the constitutive assumptions that appear in continuum mechanical theories for open lipid bilayers~\cite{tu2004geometric,biria2013continuum,guven2014terasaki}.

Our derivation gives rise to integral representations for the material parameters present in the model. Specifically, the molecular origins of the spontaneous curvatures of the edge of a lipid bilayer have been investigated. A concrete soft-core interaction potential for axisymmetric rod-like molecules was applied on the derived model to obtain those material parameters. Hence, a special form of the interaction potential suggested by Berne and Pechukas~\cite{berne1972gaussian} was employed to further explore the developed microphysical model. Assuming that the molecules are uniformly distributed along the edge, (i.e. $\Pi=$constant), the spontaneous curvatures and the torsional contribution to the energy vanish, resulting in~\eqref{nresult}, which includes only a constant part $k_{\circ}$, which can be interpreted as a line tension, and a contribution due to bending with flexural rigidity $k_1=k_2$. The dependence of the parameters $k_{\circ}$ and $k_1$ on the aspect ratio $\sigma_e/\sigma_s$ and the dimensionless cutoff distance $\delta/\sigma_\circ$, for various aspect ratios common for phospholipid molecules comprising lipid membranes, was investigated. It was concluded that increasing the cut-off distance after a value $\delta^*=3\sigma_{\circ}$ does not affect those parameters. This result and the definition $\sigma_\circ=\sqrt{2}\sigma_s$ reveals that each molecule on the edge interacts with less than $6$ molecules in its vicinity along the edge. In view of this observation, the cut-off distance was set to $\delta^*$ to study the dependence of the line tension $k_{\circ}$ and flexural rigidity $k_1$ on the molecular aspect ratio $\sigma_e/\sigma_s$. It is evident from Figure~\ref{Fig4} that the line tension $k_{\circ}$ is not as sensitive as the flexural rigidity $k_1$ to the molecular aspect ratio $\sigma_e/\sigma_s$. This difference can be quantified by considering the relative change of those quantities associated with a same increment of the molecular aspect ratio $\sigma_e/\sigma_s$. Further, with an increase in the aspect ratio, the line tension decreases while the flexural rigidity increases. This signals that for molecules with greater aspect ratios, inclusion of the bending contribution to the free-energy is of more significance. Considering that a rod with a larger diameter shows more resistance to bending, the observation that the flexural rigidity of the edge is greater for larger molecular aspect ratios (longer molecules lead to larger cross-sectional diameter of the edge), agrees with what would be expected intuitively.

The energy functional obtained in~\eqref{finalform3} was used to explore energy of the degenerate case of a circular pore on a spheroidal lipid bilayer, resulting in the expression~\eqref{example3}. The pore size needs to be greater than the mean distance of molecules $\sigma_{\circ}$. For such sizes, the contribution of the flexural part to the energy given by~\eqref{example3} is negligible, and effectively, the energy increases linearly with the pore size. Inspired by earlier investigations on rupturing of soap films~\cite{deryagin1962theory}, Litster~\cite{litster1975stability} developed a continuum model for the free-energy $\phi$ needed for opening-up of a pore in a lipid membrane in the form 
\begin{equation}
\frac{\phi}{2\pi}=\Gamma r -\frac{1}{2} \gamma_s r^2,
\label{Litster}
\end{equation} 
with $\Gamma$ the line energy of the edge and $\gamma_s$ the interfacial surface tension. It can be inferred from~\eqref{Litster} that transient pores with sizes less than a critical radius $r^*=\Gamma/\gamma_s$ tend to reseal, while those having the size exceeding this critical radius, might grow indefinitely, leading to rupture of the membrane~\cite{deserno2014fluid}. In other words, the energy to form a pore of radius $r$ is determined by a balance between two competitive contributions: the energy required to create the edge of the pore, and the energy released by the pore surface~\cite{genco1993electroporation}. Nevertheless, it is a common knowledge today (see for example the review by J\"{a}hnig~\cite{jahnig1996surface}) that lipid membranes possess zero surface tension, by which the second term on the right hand of~\eqref{Litster} vanishes. This transpires the agreement of the current result based on molecular interactions, with that obtained previously on continuum grounds. The increasing cost of generating a larger pore confirms the stability of a lipid membrane with respect to the fluctuations that might bring about transient pores. Furthermore, the growth of stable pores in homogenous lipid bilayers is only possible in the presence of external stimuli such as an electric field.

The class of the interaction potentials~\eqref{212} selected in the present study and the integration~\eqref{energy0}, only account for the lipid bilayers in which the physiochemical properties of the constituent molecules are identical. An important corollary of our model would follow from a generalization of the arguments of the interaction potential~\eqref{212} and the integration~\eqref{energy0}, to allow for the interactions between phospholipid molecules of different physiochemical properties. Such a generalization permits modeling perforated mixed lipid bilayers, such as those reported by Ogl\c{e}cka {\it et al.}~\cite{oglkecka2014oscillatory} and Jiang \& Kindt~\cite{jiang2007simulations}. Further, our model accounts for the elastic free-energy of the edge of open lipid bilayers in which the lipid molecules are tilted only at the edge, forming a semicylindrical rim along it. Another generalization of the present work would include tilt fields of smaller gradient, such as those considered by Hamm and Kozlov~\cite{hamm2000elastic}, Rangamani and Steigmann~\cite{rangamani2014variable}, and Rangamani {\it et al.}~\cite{rangamani2014small}. In such an approach, the gradual tilt at the vicinity of the edge changes the thickness of the lipid bilayer and, thus, leads to the deviation of the conformation of the edge from a semicylindrical shape. These potential generalizations remain to be investigated in future. 

\section{Acknowledgement}\label{Acknowledgement}

Financial support from National Institutes of Health (NIDCD) grant DC 005788 is gratefully acknowledged. The authors thank Mohsen Maleki for Figures~1--3.

\appendix\label{appendix}
\section{Detailed derivation of~\eqref{framede}}\label{appproof}
Let $\psi$ denote the angle between the unit normal $\bf{N}$ to the curve $\cal C$ and the unit normal $\bfn$ to the surface ${\cal S}$. The Darboux frame $\{\bft,\bfn,\bfp\}$ at any point on $\cal C$, is obtained by rotating the Frenet frame $\{\bf{t},\bf{N},\bf{B}\}$ about the unit tangent $\bf{t}$ by that angle. Hence,
\begin{equation}
\left[\begin{array}{c} {\bft} \\{\bfn} \\ {\bfp} \end{array}\right]=\begin{bmatrix}1 & 0 & 0 \\0 & \cos\psi & \sin\psi \\0 & -\sin\psi & \cos\psi \\\end{bmatrix} \left[\begin{array}{c} {\bf{t}} \\{\bf{N}} \\ {\bf{B}} \end{array} \right].
\end{equation}
Let the array ${\bf{X}=\begin{bmatrix} {\bft} &{\bfn} &{\bfp} \end{bmatrix}}^{\text{T}}$ denote the Darboux frame, and ${\bf{Y}=\begin{bmatrix} {\bf{t}} &{\bf{N}} &{\bf{B}} \end{bmatrix}}^{\text{T}}$ denote the Frenet frame at a given point on $\cal C$, where T denotes the transpose. Also let
\begin{equation}
{\bf{Q}}:=\begin{bmatrix}1 & 0 & 0 \\0 & \cos\psi & \sin\psi \\0 & -\sin\psi & \cos\psi \\\end{bmatrix}
\end{equation}
denote the transformation between the two frames. Thus, 
\begin{equation}
{\bf{X}}={\bf{Q}}{\bf{Y}}, \quad\text{or}\quad {\bf{Y}}={\bf{Q}}^{\text{T}}{\bf{X}}.
\label{xandyy}
\end{equation}
The derivative of the Frenet frame $\{\bf{t},\bf{N},\bf{B}\}$ of ${\cal C}$ with respect to the arclength $s$, follows from the Frenet-Serret formulas~\cite{kreyszig1968introduction}
\begin{equation}
\left[ \begin{array}{c}{\dot{\bf{t}}} \\{\dot{\bf{N}}} \\ {\dot{\bf{B}}} \end{array} \right]=\begin{bmatrix}0 & \kappa & 0 \\-\kappa & 0 & \tau \\0 & -\tau & 0 \\\end{bmatrix} \left[\begin{array}{c} {\bf{t}} \\{\bf{N}} \\ {\bf{B}} \end{array} \right].
\label{ydot1}
\end{equation}
Thus,
\begin{equation}
\dot{\bf{Y}}={\bf{A}}{\bf{Y}}={\bf{A}}{\bf{Q}}^{\text{T}}{\bf{X}},
\label{ydot}
\end{equation}
where
\begin{equation}
{\bf{A}}=\begin{bmatrix}0 & \kappa & 0 \\-\kappa & 0 & \tau \\0 & -\tau & 0 \\\end{bmatrix}.
\end{equation}
Taking arclength derivative from both sides of~\eqref{xandyy}$_1$ yields
\begin{equation}
\label{xdotx}
\dot{\bf{X}}=\dot{\bf{Q}}{\bf{Y}}+{\bf{Q}}\dot{\bf{Y}},
\end{equation}
where
\begin{equation}
\dot{\bf{Q}}=\begin{bmatrix}0 & 0 & 0 \\0 & -\dot{\psi}\sin\psi & \dot{\psi}\cos\psi \\0 & -\dot{\psi}\cos\psi & -\dot{\psi}\sin\psi \\\end{bmatrix}.
\end{equation}
In view of~\eqref{xandyy}$_2$ and the right-hand side of~\eqref{ydot},~\eqref{xdotx} takes the form
\begin{equation}
\dot{\bf{X}}=(\dot{\bf{Q}}{\bf{Q}}^T+{\bf{Q}}\bf{A}{\bf{Q}}^T)\,{\bf{X}},
\end{equation}
or, equivalently,
\begin{equation}\label{frenetserret2}
\left[ \begin{array}{c}{\dot{\bft}} \\{\dot{\bfn}} \\ {\dot\bfp} \end{array} \right]=\begin{bmatrix}0 & \kappa\cos\psi & -\kappa\sin\psi \\-\kappa\cos\psi & 0 & \tau+\dot\psi \\\kappa\sin\psi & -\tau-\dot\psi & 0 \\\end{bmatrix} \left[\begin{array}{c} {\bft} \\{\bfn} \\ {\bfp} \end{array} \right].
\end{equation}
In view of $\kn=\kappa\cos\psi$, $\kg=\kappa\sin\psi$, and $\tau_g=\tau+\dot\psi$,~\eqref{frenetserret2} simplifies to~\eqref{framede}.

\section{Derivation of the free-energy density{~\eqref{finalform}}}\label{Derivation of the free-energy density}

In this Appendix, the expansion of~\eqref{e11} to~\eqref{finalform} is presented. As mentioned earlier, only the molecules separated by a distance less than $\delta$ may interact. Hence, the domain of the integral with respect to $t$ in~\eqref{interaction0} is replaced by $[-\delta,\delta]$. Upon replacing $\omega(s,\theta)$ in~\eqref{e11}, and substituting $\Omega$ by its equivalent from~\eqref{212}, and applying the change of variable $t-t_\circ=s\varrho$,~\eqref{e11} takes the form
\begin{multline}
\phi=\varrho\int_{-\ell}^{\ell}\mskip-3mu\int_{\frac{-\pi}{2}}^{\frac{\pi}{2}}\mskip-5mu\int_{\frac{-\pi}{2}}^{\frac{\pi}{2}}\tilde\Omega\Big(\varrho^{-2}\bfr\cdot\bfr,\bfr\cdot\bfd(t_\circ,\theta),\\
\qquad\qquad\qquad\bfr\cdot\bfd(t_\circ+s\varrho,\eta),
\bfd(t_\circ,\theta)\cdot\bfd(t_\circ+s\varrho,\eta)\Big)\\
\Pi(t_\circ)\Pi(t_\circ+s\varrho)\, \deta \dtheta \ds.
\label{dir3}
\end{multline}
It is necessary to expand the right-hand side of~\eqref{dir3} in powers of $\varrho$ neglecting terms of $o(\varrho^2)$. Introducing the abbreviations
\begin{equation}
\left.
\begin{split}
\bfn&:=\bfn(0),\qquad \bft:=\bft(0),\qquad \bfp:=\bfp(t_{\circ}),\\[4pt]
\Pi&:=\Pi(0),\qquad\dot{\Pi}:=\dot{\Pi}(0),\quad\ddot{\Pi}:=\ddot{\Pi}(0),
\end{split}
\right\}
\label{not1}
\end{equation}  
and applying the identities given in~\eqref{framede}, the following expansions up to $\varrho^2$ are obtained:
\begin{align}
\label{ex100}
\nonumber{\bfx}(s\varrho)&={\bfx}(0)+\varrho s\,\bft+\frac{1}{2}s^2\varrho^2\kappa_n\,\bfn-\frac{1}{2}s^2\varrho^2\kappa_g\, \bfp+o(\varrho^4),\\
\nonumber\bfn(s\varrho)&=\bigg(\frac{s^2\varrho^2}{2}\big(\dot\kappa_n-\kappa_g\tau_g\big)-s\varrho\kn\bigg)\bft+\bigg(1-\frac{s^2\varrho^2}{2}\big(\kn^2+\tau^2_g\big)\bigg)\bfn\\
&\notag\qquad+\bigg(\frac{s^2\varrho^2}{2}\big(\kn\kg+\dot\tau_g\big)+s\varrho\tau_g\bigg)\bfp+o(\varrho^2),\\
\notag\bfp(s\varrho)&=\bigg(\frac{s^2\varrho^2}{2}\big(\kn\kg-\dot\tau_g\big)-s\varrho\tau_g\bigg)\bfn+\bigg(\varrho s\kg+\frac{s^2\varrho^2}{2}\big(\dot\kappa_g+\tau_g\kn\big)\bigg)\bft\\
&\qquad+\bigg(1-\frac{s^2\varrho^2}{2}\big(\tau^2_g+\kg^2\big)\bigg)\bfp+o(\varrho^2).
\end{align}
Therefore, the arguments of the interaction potential $\tilde\Omega$ in the right-hand side of~\eqref{dir3} become
\begin{align}
\notag&{{\varrho}^{-2}}{\abs{{\bfx}(t_\circ)-{\bfx}(t_\varrho)}^2}=s^2+A_1{\varrho^2}s^4+o(\varrho^2),
\\[4pt]
\notag&\big({\bfx}(t_\circ)-{\bfx}(t_\varrho)\big)\cdot\bfd(t_\circ,\theta)=A_2{\varrho^2}s^2+o(\varrho^2),
\\[4pt]
\notag&\big({\bfx}(t_\circ)-{\bfx}(t_\varrho)\big)\cdot\bfd(t_\varrho,\eta)=A_3{\varrho^2}s^2+o(\varrho^2),
\\[4pt]
&\bfd(t_\circ,\theta)\cdot\bfd(t_\varrho,\eta)=A_4+A_5{\varrho} s+A_6{\varrho^2} s^2+o(\varrho^2),
\end{align}  
where $t_\varrho=t_\circ+s\varrho$, and
\begin{align}
\nonumber&A_1=-\frac{\kappa^2}{12},
\quad 
A_2=\frac{\kg\ct-\kn\st}{2},\quad A_3=\frac{\kn\se-\kg\ce}{2},
\\
\notag&A_4=\cos(\theta-\eta)\quad A_5=-\tau_g\sin(\theta-\eta),
\\[4pt]
\notag&A_6=\frac{\kn\kg}{2}\sin(\theta+\eta)-\frac{\dot{\tau_g}}{2}\sin(\theta-\eta)-\frac{\tau^2_g}{2}\cos(\theta-\eta)\\
&\qquad-\frac{\kg^2}{2}\ct\,\ce-\frac{\kn^2}{2}\st\,\se.
\label{dir5}
\end{align}
Expanding $\tilde\Omega$ up to $\varrho^2$ and using
\begin{equation}
\Pi(s\varrho)=\Pi+s\varrho\dot{\Pi}+\frac{s^2\varrho^2}{2}\ddot{\Pi}+o(\varrho^2),
\end{equation}
results in the following energy-density for the edge
%
\begin{align}
\nonumber\phi&=\int_{-\ell}^{\ell}\int_{\frac{-\pi}{2}}^{\frac{\pi}{2}}\int_{\frac{-\pi}{2}}^{\frac{\pi}{2}}\bigg\{\varrho\,\bar\Omega(s,\theta,\eta)\Pi^2 \deta \dtheta \ds+\frac{1}{2}\varrho^3\,\bar\Omega(s,\theta,\eta)s^2\Pi\ddot{\Pi}\bigg\} \deta \dtheta \ds\\
&\nonumber+\kappa^2_g\bigg\{\varrho^3\int_{-\ell}^{\ell}\int_{\frac{-\pi}{2}}^{\frac{\pi}{2}}\mskip-3mu\int_{\frac{-\pi}{2}}^{\frac{\pi}{2}}\frac{-s^2\Pi^2}{12}\Big(\bar\Omega_1(s,\theta,\eta)s^2+6\,\bar\Omega_4(s,\theta,\eta)\ct\ce\Big)\, \deta \dtheta \ds\bigg\}\\
&\nonumber+\kappa^2_n\bigg\{\varrho^3\int_{-\ell}^{\ell}\mskip-3mu\int_{\frac{-\pi}{2}}^{\frac{\pi}{2}}\mskip-3mu\int_{\frac{-\pi}{2}}^{\frac{\pi}{2}}\frac{-s^2\Pi^2}{12}\Big(\bar\Omega_1(s,\theta,\eta)s^2+6\,\bar\Omega_4(s,\theta,\eta)\st\se\Big)\, \deta \dtheta \ds\bigg\}\\
&\nonumber+\kappa_g\bigg\{\varrho^3\int_{-\ell}^{\ell}\mskip-3mu\int_{\frac{-\pi}{2}}^{\frac{\pi}{2}}\mskip-3mu\int_{\frac{-\pi}{2}}^{\frac{\pi}{2}}\frac{s^2\Pi^2}{2}\Big(\bar\Omega_2(s,\theta,\eta)\ct-\bar\Omega_3(s,\theta,\eta)\ce\Big)\, \deta \dtheta \ds\bigg\}\\
&\nonumber+\kappa_n\bigg\{\varrho^3\int_{-\ell}^{\ell}\mskip-3mu\int_{\frac{-\pi}{2}}^{\frac{\pi}{2}}\mskip-3mu\int_{\frac{-\pi}{2}}^{\frac{\pi}{2}}\frac{s^2\Pi^2}{2}\Big(\bar\Omega_3(s,\theta,\eta)\se-\bar\Omega_2(s,\theta,\eta)\st\Big)\, \deta \dtheta \ds\bigg\}\\
&\nonumber+\tau^2_g\bigg\{\varrho^3\int_{-\ell}^{\ell}\mskip-3mu\int_{\frac{-\pi}{2}}^{\frac{\pi}{2}}\mskip-3mu\int_{\frac{-\pi}{2}}^{\frac{\pi}{2}}\frac{s^2\Pi^2}{2}\Big(\bar\Omega_{44}(s,\theta,\eta)\sin^2(\theta-\eta)-\bar\Omega_4(s,\theta,\eta)\cos(\theta-\eta)\Big)\deta \dtheta \ds\bigg\}\\[4pt]
&+\kappa_n\kappa_g\bigg\{\varrho^3\int_{-\ell}^{\ell}\mskip-3mu\int_{\frac{-\pi}{2}}^{\frac{\pi}{2}}\mskip-3mu\int_{\frac{-\pi}{2}}^{\frac{\pi}{2}}\frac{s^2\Pi^2}{2}\bar\Omega_4(s,\theta,\eta)\sin(\theta+\eta)\, \deta \dtheta \ds\bigg\} ,
\label{proof}
\end{align}
%
where
\begin{equation}
\left.
\begin{split}
&\bar\Omega(s,\theta,\eta):=\tilde\Omega(s^2,0,0,\cos(\theta-\eta)),
\\[4pt]
&\bar\Omega_i(s,\theta,\eta):=\tilde\Omega_{,i}(s^2,0,0,\cos(\theta-\eta)), 
\\[4pt]
&\bar\Omega_{ii}(s,\theta,\eta):=\tilde\Omega,_{ii}(s^2,0,0,\cos(\theta-\eta)),
\end{split}
\right\} \quad i\in\{1,2,3,4\}.
\label{dir52}
\end{equation}
Note that $\bar\Omega$, $\bar\Omega_i$ and $\bar\Omega_{ii}$ vanish when ${s}>\ell$. Thus, the final expression for the free-energy density of the edge becomes
\begin{equation}
\phi=k_\circ+k_1\kg^2+k_2\kn^2+k_3\kg+k_4\kn\\
+k_5\kn\kg+k_6\tau_g^2,
\label{dir7}
\end{equation}
or, equivalently,
\begin{equation}
\phi=k_\circ+k_1(\kg-{\kappa}_{g\circ})^2+k_2(\kn-{\kappa}_{n\circ})^2+k_5\kn\kg+k_6\tau_g^2,
\label{dir7}
\end{equation}
where the parameters $k_\circ$--$k_6$ are given by the integral representations in~\eqref{proof}.

\section{Integral representations $I_\circ$, $I$ and $J$ in~\eqref{coefficients1}:}\label{Appendix2}

\begin{align}
&\nonumber I_\circ=\int_{\frac{-\pi}{2}}^{\frac{\pi}{2}}\mskip-6mu\int_{\frac{-\pi}{2}}^{\frac{\pi}{2}}{\sqrt{1-\chi^2\cos^2(\theta-\eta)}} \deta \dtheta,
\\[4pt]
&\notag I=\int_{\frac{-\pi}{2}}^{\frac{\pi}{2}}\mskip-6mu\int_{\frac{-\pi}{2}}^{\frac{\pi}{2}}\frac{1}{\sqrt{1-\chi^2\cos^2(\theta-\eta)}} \deta \dtheta,
\\[4pt]& J=\int_{\frac{-\pi}{2}}^{\frac{\pi}{2}}\mskip-6mu\int_{\frac{-\pi}{2}}^{\frac{\pi}{2}}\frac{\cos(2\theta-2\eta)}{\sqrt{1-\chi^2\cos^2(\theta-\eta)}} \deta \dtheta.
\label{integrals}
\end{align}
%

%
\bibliography{mybibfile}

\begin{thebibliography}{10}
\expandafter\ifx\csname url\endcsname\relax
  \def\url#1{\texttt{#1}}\fi
\expandafter\ifx\csname urlprefix\endcsname\relax\def\urlprefix{URL }\fi
\expandafter\ifx\csname href\endcsname\relax
  \def\href#1#2{#2} \def\path#1{#1}\fi

\bibitem{israelachvili2011intermolecular}
J.~N. Israelachvili, Intermolecular and surface forces: revised third edition,
  Academic press, 2011.

\bibitem{israelachvili1977theory}
J.~N. Israelachvili, D.~J. Mitchell, B.~W. Ninham, Theory of self-assembly of
  lipid bilayers and vesicles, Biochimica et Biophysica Acta (BBA)-Biomembranes
  470~(2) (1977) 185--201.

\bibitem{israelachvili1976theory}
J.~N. Israelachvili, D.~J. Mitchell, B.~W. Ninham, Theory of self-assembly of
  hydrocarbon amphiphiles into micelles and bilayers, Journal of the Chemical
  Society, Faraday Transactions 2: Molecular and Chemical Physics 72 (1976)
  1525--1568.

\bibitem{lodish2000molecular}
H.~F. Lodish, A.~Berk, S.~L. Zipursky, P.~Matsudaira, D.~Baltimore, J.~Darnell,
  et~al., Molecular cell biology, Vol.~4, WH Freeman New York, 2000.

\bibitem{saitoh1998opening}
A.~Saitoh, K.~Takiguchi, Y.~Tanaka, H.~Hotani, Opening-up of liposomal
  membranes by talin, Proceedings of the National Academy of Sciences 95~(3)
  (1998) 1026--1031.

\bibitem{luckey2014membrane}
M.~Luckey, Membrane structural biology: with biochemical and biophysical
  foundations, Cambridge University Press, 2014.

\bibitem{seifert1997configurations}
U.~Seifert, Configurations of fluid membranes and vesicles, Advances in Physics
  46~(1) (1997) 13--137.

\bibitem{lasic1988mechanism}
D.~D. Lasic, The mechanism of vesicle formation., Biochemical Journal 256~(1)
  (1988) 1.

\bibitem{tsong1991electroporation}
T.~Y. Tsong, Electroporation of cell membranes, Biophysical Journal 60~(2)
  (1991) 297--306.

\bibitem{marmottant2003controlled}
P.~Marmottant, S.~Hilgenfeldt, Controlled vesicle deformation and lysis by
  single oscillating bubbles, Nature 423~(6936) (2003) 153--156.

\bibitem{fromherz1986discoid}
P.~Fromherz, C.~R{\"o}cker, D.~R{\"u}ppel, From discoid micelles to spherical
  vesicles. the concept of edge activity, Faraday Discussions of the Chemical
  Society 81 (1986) 39--48.

\bibitem{tu2004geometric}
Z.~Tu, Z.~Ou-Yang, A geometric theory on the elasticity of bio-membranes,
  Journal of Physics A: Mathematical and General 37~(47) (2004) 11407.

\bibitem{tu2008elastic}
Z.~Tu, Z.~Ou-Yang, Elastic theory of low-dimensional continua and its
  applications in bio-and nano-structures, Journal of Computational and
  Theoretical Nanoscience 5~(4) (2008) 422--448.

\bibitem{boal1992scaling}
D.~H. Boal, M.~Rao, Scaling behavior of fluid membranes in three dimensions,
  Physical Review A 45~(10) (1992) R6947.

\bibitem{capovilla2002lipid}
R.~Capovilla, J.~Guven, J.~Santiago, Lipid membranes with an edge, Physical
  Review E 66~(2) (2002) 021607.

\bibitem{tu2010compatibility}
Z.~Tu, Compatibility between shape equation and boundary conditions of lipid
  membranes with free edges, The Journal of Chemical Physics 132~(8) (2010)
  084111.

\bibitem{tu2011geometry}
Z.~Tu, Geometry of membranes, arXiv preprint arXiv:1106.2370.

\bibitem{tu2003lipid}
Z.~Tu, Z.~Ou-Yang, Lipid membranes with free edges, Physical Review E 68~(6)
  (2003) 061915.

\bibitem{may2000molecular}
S.~May, A molecular model for the line tension of lipid membranes, The European
  Physical Journal E 3~(1) (2000) 37--44.

\bibitem{keller1991flexural}
J.~B. Keller, G.~J. Merchant, Flexural rigidity of a liquid surface, Journal of
  Statistical Physics 63~(5-6) (1991) 1039--1051.

\bibitem{seguin2014microphysical}
B.~Seguin, E.~Fried, Microphysical derivation of the canham--helfrich
  free-energy density, Journal of Mathematical Biology 68~(3) (2014) 647--665.

\bibitem{asgari2015elastic}
M.~Asgari, Elastic free-energy of wormlike micellar chains: theory and
  suggested experiments, arXiv preprint arXiv:1502.02338.

\bibitem{jiang2004molecular}
F.~Y. Jiang, Y.~Bouret, J.~T. Kindt, Molecular dynamics simulations of the
  lipid bilayer edge, Biophysical Journal 87~(1) (2004) 182--192.

\bibitem{wohlert2006free}
J.~Wohlert, W.~Den~Otter, O.~Edholm, W.~Briels, Free energy of a trans-membrane
  pore calculated from atomistic molecular dynamics simulations, The Journal of
  Chemical Physics 124~(15) (2006) 154905.

\bibitem{de2006coarse}
J.~de~Joannis, F.~Y. Jiang, J.~T. Kindt, Coarse-grained model simulations of
  mixed-lipid systems: composition and line tension of a stabilized bilayer
  edge, Langmuir 22~(3) (2006) 998--1005.

\bibitem{karatekin2003cascades}
E.~Karatekin, O.~Sandre, H.~Guitouni, N.~Borghi, P.-H. Puech,
  F.~Brochard-Wyart, Cascades of transient pores in giant vesicles: line
  tension and transport, Biophysical Journal 84~(3) (2003) 1734--1749.

\bibitem{berne1972gaussian}
B.~J. Berne, P.~Pechukas, Gaussian model potentials for molecular interactions,
  The Journal of Chemical Physics 56~(8) (1972) 4213--4216.

\bibitem{gay1981modification}
J.~Gay, B.~Berne, Modification of the overlap potential to mimic a linear
  site--site potential, The Journal of Chemical Physics 74~(6) (1981)
  3316--3319.

\bibitem{guggenheimer1977differential}
H.~Guggenheimer, Differential geometry, 1963 (1977).

\bibitem{synge1969tensor}
J.~L. Synge, Tensor calculus, Vol.~5, Courier Corporation, 1969.

\bibitem{guven2014terasaki}
J.~Guven, G.~Huber, D.~M. Valencia, Terasaki spiral ramps in the rough
  endoplasmic reticulum, Physical Review Letters 113~(18) (2014) 188101.

\bibitem{zhelev1993tension}
D.~V. Zhelev, D.~Needham, Tension-stabilized pores in giant vesicles:
  determination of pore size and pore line tension, Biochimica et Biophysica
  Acta (BBA)-Biomembranes 1147~(1) (1993) 89--104.

\bibitem{truesdell2004non}
C.~Truesdell, W.~Noll, The non-linear field theories of mechanics, Springer,
  2004.

\bibitem{whitehead2001molecular}
L.~Whitehead, C.~M. Edge, J.~W. Essex, Molecular dynamics simulation of the
  hydrocarbon region of a biomembrane using a reduced representation model,
  Journal of Computational Chemistry 22~(14) (2001) 1622--1633.

\bibitem{mashaghi2012hydration}
A.~Mashaghi, P.~Partovi-Azar, T.~Jadidi, N.~Nafari, P.~Maass, M.~R.~R. Tabar,
  M.~Bonn, H.~J. Bakker, Hydration strongly affects the molecular and
  electronic structure of membrane phospholipids, The Journal of Chemical
  Physics 136~(11) (2012) 114709.

\bibitem{jiang2007simulations}
Y.~Jiang, J.~T. Kindt, Simulations of edge behavior in a mixed-lipid bilayer:
  fluctuation analysis, The Journal of Chemical Physics 126~(4) (2007) 045105.

\bibitem{pera2015edge}
H.~Pera, J.~Kleijn, F.~Leermakers, On the edge energy of lipid membranes and
  the thermodynamic stability of pores, The Journal of Chemical Physics 142~(3)
  (2015) 034101.

\bibitem{biria2013continuum}
A.~Biria, M.~Maleki, E.~Fried, Continuum theory for the edge of an open lipid
  bilayer, Advances in Applied Mechanics 21 (2013) 1--78.

\bibitem{deryagin1962theory}
B.~Deryagin, Y.~V. Gutop, Theory of the breakdown (rupture) of free films,
  Kolloidn. Zh 24 (1962) 370--374.

\bibitem{litster1975stability}
J.~Litster, Stability of lipid bilayers and red blood cell membranes, Physics
  Letters A 53~(3) (1975) 193--194.

\bibitem{deserno2014fluid}
M.~Deserno, Fluid lipid membranes: From differential geometry to curvature
  stresses, Chemistry and Physics of Lipids.

\bibitem{genco1993electroporation}
I.~Genco, A.~Gliozzi, A.~Relini, M.~Robello, E.~Scalas, Electroporation in
  symmetric and asymmetric membranes, Biochimica et Biophysica Acta
  (BBA)-Biomembranes 1149~(1) (1993) 10--18.

\bibitem{jahnig1996surface}
F.~J{\"a}hnig, What is the surface tension of a lipid bilayer membrane?,
  Biophysical Journal 71~(3) (1996) 1348.

\bibitem{oglkecka2014oscillatory}
K.~Ogl{\c{e}}cka, P.~Rangamani, B.~Liedberg, R.~S. Kraut, A.~N. Parikh,
  Oscillatory phase separation in giant lipid vesicles induced by transmembrane
  osmotic differentials, eLife 3 (2014) e03695.

\bibitem{hamm2000elastic}
M.~Hamm, M.~Kozlov, Elastic energy of tilt and bending of fluid membranes, The
  European Physical Journal E 3~(4) (2000) 323--335.

\bibitem{rangamani2014variable}
P.~Rangamani, D.~Steigmann, Variable tilt on lipid membranes, Proceedings of
  the Royal Society A: Mathematical, Physical and Engineering Science
  470~(2172) (2014) 20140463.

\bibitem{rangamani2014small}
P.~Rangamani, A.~Benjamini, A.~Agrawal, B.~Smit, D.~J. Steigmann, G.~Oster,
  Small scale membrane mechanics, Biomechanics and modeling in mechanobiology
  13~(4) (2014) 697--711.

\bibitem{kreyszig1968introduction}
E.~Kreyszig, Introduction to differential geometry and Riemannian geometry,
  Vol.~16, University of Toronto Press, 1968.

\end{thebibliography}
\bibliographystyle{elsarticle-num}

\end{document}